\documentclass[10pt, aps, prl, twocolumn, superscriptaddress, floatfix, showpacs, longbibliography]{revtex4-1}

\usepackage{graphicx}
\usepackage{amsfonts}
\usepackage{amssymb}
\usepackage{amsmath}
\usepackage{txfonts}
\usepackage{lipsum}
\usepackage{color}
\usepackage{wasysym}
\usepackage{hyperref}
\usepackage{bbold}
\usepackage{comment}
\usepackage{etoolbox}
\graphicspath{{figures/}}
\usepackage[normalem]{ulem} 
\usepackage{mathtools}

\usepackage{mathrsfs}

\usepackage{units}

\newcommand{\qv}{\mathbf{q}}
\newcommand{\kv}{\mathbf{k}}

\usepackage{letltxmacro}
\LetLtxMacro{\oldsqrt}{\sqrt}
\renewcommand{\sqrt}[2][\mkern8mu]{\mkern-6mu\mathop{}\oldsqrt[#1]{#2}}

\begin{document}

\title{
Doping-dependent charge- and spin-density wave orderings \\ in a monolayer of Pb adatoms on Si(111)
}

\author{M. Vandelli}
\affiliation{Institut f{\"u}r Theoretische Physik, Universit{\"a}t Hamburg, Notkestra{\ss}e 9 , 22607 Hamburg, Germany}
\affiliation{The Hamburg Centre for Ultrafast Imaging, Luruper Chaussee 149, 22761 Hamburg, Germany}
\affiliation{Max Planck Institute for the Structure and Dynamics of Matter,
Center for Free Electron Laser Science, 22761 Hamburg, Germany}

\author {A. Galler}
\affiliation{Max Planck Institute for the Structure and Dynamics of Matter, Center for Free Electron Laser Science, 22761 Hamburg, Germany}

\author{A. Rubio}
\affiliation{Max Planck Institute for the Structure and Dynamics of Matter, Center for Free Electron Laser Science, 22761 Hamburg, Germany}
\affiliation{Center for Computational Quantum Physics, Flatiron Institute, 162 5th Avenue, New York, NY 10010, USA}
\affiliation{Nano-Bio Spectroscopy Group and ETSF, Universidad del Pa\'is Vasco, 20018 San Sebast\'ian, Spain}

\author{A. I. Lichtenstein}
\affiliation{Institut f{\"u}r Theoretische Physik, Universit{\"a}t Hamburg, Notkestra{\ss}e 9 , 22607 Hamburg, Germany}
\affiliation{The Hamburg Centre for Ultrafast Imaging, Luruper Chaussee 149, 22761 Hamburg, Germany}

\author{S. Biermann}
\affiliation{CPHT, CNRS, {\'E}cole polytechnique, Institut Polytechnique de Paris, 91120 Palaiseau, France}
\affiliation{Coll\`ege de France, 11 place Marcelin Berthelot, 75005 Paris, France}
\affiliation{Department of Physics, Division of Mathematical Physics, Lund University, Professorsgatan 1, 22363 Lund, Sweden}
\affiliation{European Theoretical Spectroscopy Facility, 91128 Palaiseau, France}

\author{E. A. Stepanov}
\affiliation{CPHT, CNRS, {\'E}cole polytechnique, Institut Polytechnique de Paris, 91120 Palaiseau, France}

\begin{abstract}
In this work we computed the phase diagram as a function of temperature and doping for a system of lead adatoms allocated periodically on a silicon (111) surface. 
This Si(111):Pb material is characterized by a strong and long-ranged Coulomb interaction, a relatively large value of the spin-orbit coupling, and a structural phase transition that occurs at low temperature. In order to describe the collective electronic behavior in the system, we perform many-body calculations consistently taking all these important features into account.
We find that charge- and spin-density wave orderings coexist with each other in several regions of the phase diagram.
This result is in agreement with the recent experimental observation of a chiral spin texture in the charge density wave phase in this material.
We also find that geometries of the charge and spin textures strongly depend on the doping level. 
The formation of such a rich phase diagram in the Si(111):Pb material can be explained by a combined effect of the lattice distortion and electronic correlations.
\end{abstract}

\maketitle

\section{Introduction}

Recent advances in scanning tunneling microscopy have enabled an extensive control of single atoms placed on different surfaces~\cite{Custance2009, Khajetoorians2010, doi:10.1126/science.1201725, Khajetoorians2012, PhysRevLett.110.126804, PhysRevLett.111.157204, doi:10.1126/science.aad8038, Khajetoorians2016}. 
These techniques paved the way for creation and investigation of a new class of synthetic two-dimensional materials constituted by atomic structures allocated on top of different substrates, e.g., systems of tin (Sn) and lead (Pb) adatoms disposed periodically on silicon Si(111)~\cite{PhysRevB.60.13328, PhysRevB.68.235332, PhysRevB.71.033403, PhysRevLett.98.126401, Zhang2010, Li2013, PhysRevLett.120.196402}, germanium Ge(111)~\cite{PhysRevLett.79.2859, PhysRevB.64.075405, PhysRevB.104.045126}, or SiC(0001)~\cite{PhysRevLett.114.247602} surfaces. 
The possibility of tuning the structure and the chemical composition in these  two-dimensional systems allows for a direct modification of properties  in a similar way to cold atoms or Moir\'e heterostructures as proposed in Ref.~\cite{Kennes2021}.
As a consequence, they can be seen as a promising platform for simulating various quantum effects~\cite{Polini2013, Khajetoorians2019, PhysRevLett.128.167002}.

At the band structure level, depositing a monolayer of group-IV atoms on a Si(111), Ge(111), or SiC(0001) surface leads to the formation of a half-filled narrow band that is well-separated from the bands of an insulating background.
On the one hand, this situation could allow for an application of the most advanced theoretical many-body approaches developed to date for model single-band systems (see, e.g., Refs.~\cite{RevModPhys.90.025003, PhysRevX.11.011058}).
On the other hand, these materials exhibit a number of non-trivial features that make the solution of the problem not straightforward.
For instance, the wave function of single-particle states is very extended, which results in a strong and long-ranged Coulomb interaction~\cite{Hansmann_2013, PhysRevLett.110.166401} that has to be taken into account. 
Another important aspect that has to be considered is the strong spin-orbit coupling (SOC) that emerges in the case of heavy adsorbants (Sn, Pb, etc.)~\cite{PhysRevB.94.224418}.

The interplay between collective excitations and structural effects has been extensively investigated using a combination of experimental and theoretical methods in two-dimensional materials~\cite{Lee2021,Antonelli2022,Lahneman2022,Luo2022}.
However, until recently, the theoretical investigation of these surface nanostructures was mostly dedicated to the description of the metal-insulator transitions observed in scanning tunneling spectroscopy and photoemission spectroscopy experiments~\cite{PhysRevLett.98.086401, PhysRevLett.110.166401, Li2013}. 
Much less attention has been paid to collective electronic effects and, in particular, to magnetic properties, and the obtained results were controversial~\cite{Hansmann2016}.
First-principles simulations using density functional theory predicted an antiferromagnetic ground state for the Si(111):Sn material~\cite{PhysRevB.82.035116}. 
On the other hand, it has been shown that taking into account more distant hopping processes instead stabilises a row-wise collinear order~\cite{PhysRevB.83.041104}. 
It should also be noted, that both these calculations were performed without considering the effect of SOC, which may substantially affect the magnetic state.
Unfortunately, there is still no direct experimental confirmation of which magnetic ordering is actually realised in the material.  

Theoretically, the phase diagram of Pb adatoms deposited on a Si(111) surface is one of the most poorly-understood features in this class of compounds. Similarly to the Si(111):Sn material, the Bravais lattice of the Pb adatom system is rotated by $30^{\circ}$ with
respect to the substrate. A known peculiarity of the triangular lattice is a high degree of frustration that can lead to a non-trivial competition between different ordering phenomena. 
In addition, Si(111):Pb displays very strong on-site and spatial electron-electron interactions~\cite{PhysRevLett.110.166401}, which makes the system an ideal candidate to study charge and spin fluctuations.
Finally, one would also expect noticeable effects related to the SOC, since the Pb adatoms have a sufficiently large atomic number~\cite{PhysRevB.94.224418}. 
In particular, the SOC results in a splitting of the Fermi surface, which can be observed experimentally in the quasiparticle interference pattern. Additionally, the SOC gives rise to the magnetic Dzyaloshinskii–Moriya interaction, which, in turn, can lead to the formation of chiral spin textures with non-commensurate ordering vectors~\cite{PhysRevB.94.224418}. Previous calculations on the phases of this system made use of methods unable to properly account for the short and long-range correlations appearing in this system, such as DFT~\cite{PhysRevLett.120.196402}, Hartree-Fock~\cite{PhysRevB.94.224418} and cluster methods~\cite{PhysRevLett.123.086401}, or approaches that do not include magnetic fluctuations~\cite{Hansmann_2013}. Notably, DFT predicts a metallic behavior~\cite{PhysRevLett.120.196402}, whereas more correlated methods converge to a Mott insulating behavior~\cite{PhysRevB.94.224418, Hansmann_2013}.

Experimentally, it was observed that the Si(111):Pb system indeed shows a non-trivial behavior related to the above-mentioned features.
Several different arrangements of the atoms on the surface were identified, namely a ${\sqrt{3}\times\sqrt{3}}$ phase with respect to the underlying Si surface, a ${3\times3}$ phase, and a ${\sqrt{7}\times\sqrt{3}}$ phase. Recent findings indicate the presence of superconductivity in the ${\sqrt{7}\times\sqrt{3}}$ at low temperatures~\cite{Brun2014, Brun_2016} as well as chiral superconductivity in the Si(111):Sn system~\cite{Ming2023}. Unknown superconductive phases could appear also in other surface reconstructions of Si(111):Pb, likely coexisting with magnetic phases or CDW. The lattice distortion would likely play a crucial role as well. Due to the similarities with Si(111):Sn, we would expect the chirality to be still present. In addition to that, the strong spin-orbit coupling could even lead to more exotic forms of superconductivity~\cite{PhysRevLett.87.037004}.

There exists a compelling evidence that the system with $1/3$ coverage exhibits a structural transition to a ${3\times 3}$ charge density wave (CDW) phase at a temperature of 86~K~\cite{PhysRevLett.94.046101, PhysRevB.75.155411, PhysRevB.107.035125}. 
It is still a matter of on-going research to understand whether this transition has to be attributed to a Peierls-like mechanism, an intrinsic asymmetry induced by the interaction with the substrate, or to strong electronic correlations, as claimed in Ref.~\cite{PhysRevLett.123.086401}.
Remarkably, a similar transition takes place in Ge(111):Pb and Ge(111):Sn~\cite{Carpinelli1996, PhysRevLett.79.2859}, but not in the Si(111):Sn compound. 
In addition, by using scanning tunnelling microscopy (STM) it has been found that the quasi-particle interference patterns are influenced by the strong value of the SOC giving rise to a chiral spin structure at low temperatures inside the CDW phase~\cite{PhysRevLett.120.196402, PhysRevLett.123.086401}.

The experimental study of the low-temperature ${3\times 3}$ phase of Si(111):Pb is technically non-trivial~\cite{PhysRevB.107.035125}. 
This phase is difficult to grow as an extended phase limiting the experimental probes that can be used.
For this reason, the investigation of this system has so far been limited to STM experiments. 
In order to perform STM measurements, it is necessary to induce a finite conductance in the system. 
To this aim, slightly doped substrates have to be used~\cite{PhysRevLett.120.196402, PhysRevLett.123.086401}. Depending on the doping level of the substrate, adatoms can exert an attractive or repulsive force on the impurities in the bulk, that can strongly affect the doping level of the surface band~\cite{PhysRevLett.119.266802, PhysRevLett.125.117001}. As a consequence, the use of Si substrates with strong electron-doping~\cite{PhysRevLett.120.196402_sm} or hole-doping~\cite{PhysRevLett.123.086401_sm} can induce a significant doping on the surface states.
Additional data with accurate experimental control over doping conditions would be crucial to shed a light on the observed phases, since the doping in the system may strongly affect collective electronic effects and related phases. 
The effect of doping could also explain a crucial difference between theoretical results and experiments. Calculations with correlated theories predict a Mott insulating behavior, while the measured STM spectrum is metallic~\cite{PhysRevLett.120.196402, PhysRevLett.123.086401}. This apparent contradiction may be explained by noting that in a Mott insulator an arbitrarily small doping level can induce a metallic behavior.
For this reason, a careful investigation of the temperature \emph{vs} doping phase diagram is absolutely necessary to explain the experimentally observed effects in Si(111):Pb. 

In this work, we use advanced many-body techniques to analyse collective electronic effects in Si(111):Pb as a function of temperature and doping. 
We find a very rich phase diagram comprising charge- and spin-density wave phases characterised by different ordering vectors. 
By comparing results for the ${\sqrt{3} \times \sqrt{3}}$ and ${3 \times 3}$ structures, we find that different CDW orderings can originate from either a structural transition due to an asymmetric interaction of adatoms with the substrate, or from strong electronic correlations depending on the doping level. 
Further, we observe that the spin ordering in the system also depends on the doping.
These results illustrate that varying the doping level in the Si(111):Pb material represents an efficient way of switching between different CDW and magnetic phases. 
In addition, we argue that a simultaneous detection of the charge- and spin-density orderings in an experiment can help to understand in which part of the complex  temperature \emph{vs} doping phase diagram the measured system is located.

\section{Results}

\begin{figure}[b!]
    \includegraphics[width=\linewidth]{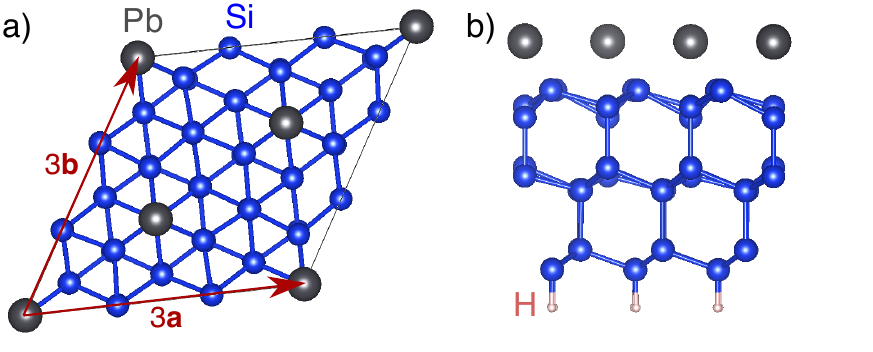} 
    \caption{(a) Top view of the Si(111):Pb surface reconstruction. Shown is one ${3\times3}$ unit cell. (b) Side view of the slab geometry adopted in the DFT structural relaxations consisting of 1/3 monolayer of Pb adatoms on
    top of three Si bi-layers, bottom terminated by hydrogen capping. 
    The uniform distance between the Pb adatoms and the substrate corresponds to a high-temperature ${\sqrt{3} \times \sqrt{3}}$ phase. In the low-temperature ${3\times3}$ reconstruction the Pb adatoms form a ``1-up-2-down'' configuration with respect to the substrate.}
    \label{fig:struct}
\end{figure}

\paragraph{\bf Model.} 

According to density functional theory (DFT) calculations, the Si(111):Pb system with 1/3 coverage in the high-temperature ${\sqrt{3} \times \sqrt{3}}$ phase (Fig.~\ref{fig:struct}) exhibits a narrow half-filled band at the Fermi level, well separated from the rest of the bands~\cite{Hansmann_2013, PhysRevLett.110.166401, PhysRevB.94.224418}.
In the maximally localized Wannier basis, this band has a $p_{z}$ character, and the corresponding Wannier orbitals are centered at the Pb adatom sites.
We thus employ the following single-band interacting electronic model derived from the first principle DFT calculations:
\begin{align}
\hat{H} &= \sum_{ij,\,\sigma\sigma'} c^{\dagger}_{i\sigma} \left(t^{\phantom{\dagger}}_{ij} \, \delta^{\phantom{\dagger}}_{\sigma\sigma'} + i\,\boldsymbol{\gamma}^{\phantom{\dagger}}_{ij} \cdot \boldsymbol{\sigma}^{\phantom{\dagger}}_{\sigma\sigma'} \right) c^{\phantom{\dagger}}_{j\sigma'}
+ {\sum_{i} \Delta_{i}n_{i}} \notag\\
&+ U^{\phantom{\dagger}}\sum_{i} \, n^{\phantom{\dagger}}_{i\uparrow} n^{\phantom{\dagger}}_{i\downarrow} 
+ \frac12\sum_{{i\neq j},} V^{\phantom{\dagger}}_{{ij}} \, n^{\phantom{\dagger}}_{i} n^{\phantom{\dagger}}_{j} + \frac12\sum_{{i\neq j}} J^{\phantom{\dagger}}_{{ij}} \, {\bf S}^{\phantom{\dagger}}_{i} \cdot {\bf S}^{\phantom{\dagger}}_{j}\,.
\label{eq_hamilt}
\end{align}
In this expression, $c^{(\dagger)}_{i\sigma}$ corresponds to an annihilation (creation) operator for an electron on the lattice site $i$ with the spin projection ${\sigma\in\{\uparrow,\downarrow\}}$. 
$t_{ij}$ corresponds to the hopping amplitude between $i$ and $j$ lattice sites, while $\Delta_{i}$ indicates the local on-site potentials.
The considered Hamiltonian accounts for the SOC in the Rashba form~\cite{Rashba, PhysRevB.52.10239} of a spin-dependent imaginary hopping ${\boldsymbol{\gamma}^{\phantom{\dagger}}_{ij} = \gamma_{|i-j|} \, \left(\hat{r}_{ij} \times \hat{z} \right)}$. 
The Coulomb interaction between electronic densities ${n_i = \sum_{\sigma}n_{i\sigma}}$, where ${n_{i\sigma} = c^{\dagger}_{i\sigma}c^{\phantom{\dagger}}_{i\sigma}}$, is explicitly divided into the local $U$ and the non-local $V_{ij}$ parts. 
$J_{ij}$ represents the direct ferromagnetic exchange interaction between the magnetic densities
${{\bf S}_i = \sum_{\sigma\sigma'}c^{\dagger}_{i\sigma}\boldsymbol{\sigma}^{\phantom{\dagger}}_{\sigma\sigma'}c^{\phantom{\dagger}}_{i\sigma'}}$, where ${\boldsymbol{\sigma} = \{\sigma^{x}, \sigma^{y}, \sigma^{z}\}}$ is a vector of Pauli matrices. 

In momentum-space, one can write the Fourier transform of the hopping amplitudes as ${\varepsilon^{\sigma\sigma'}_{\kv,ll'} = t_{\kv,ll'}\delta_{\sigma\sigma'} + i\,\vec{\gamma}_{\kv,ll'}\cdot\vec{\sigma}_{\sigma\sigma'}}$ with $l^{(\prime)}$ labeling nonequivalent lattice sites within the unit cell.
Further, we focus on the two distinct structures of the Si(111):Pb material.
In the high-temperature ${\sqrt{3}\times\sqrt{3}}$ structure the Pb adatoms form a triangular lattice with identical lattice sites, so we set ${l=l'}$. Upon decreasing the temperature, the system undergoes a structural transition, which results in a ${3\times 3}$ reconstruction of the adatoms. 
The resulting structure has the form of an effective triangular lattice, but the unit cell contains three Pb atoms. 
Lattice relaxations within the generalized gradient approximation (GGA) and experiments show that these three Pb atoms  display a corrugated ``1-up-2-down'' configuration with respect to a flat surface~\cite{PhysRevLett.94.046101, PhysRevB.75.155411, PhysRevB.107.035125}. 
We find that a local potential $\Delta_l$ with ${l\in\{1,2,3\}}$ is sufficient to describe the position of non-equivalent sites within the unit cell.
This potential is set to zero in the ${\sqrt{3}\times\sqrt{3}}$ structure, while it is non-zero in the $3\times 3$ structure because of the substrate-induced deformation, which corresponds to a static electron-phonon interaction~\cite{PhysRevB.104.045126}. 
In this regard, the high-temperature ${\sqrt{3}\times\sqrt{3}}$ phase can be seen as a time-averaged ${3\times3}$ structure, due to dynamical fluctuations of the adatom height~\cite{PhysRevLett.82.442, PhysRevB.70.155334}.
The values of all model parameters and details of the DFT calculations are given in the Methods section. 

\paragraph{\bf Detection of collective electronic instabilities.}

Instabilities related to collective electronic fluctuations in the charge ($c$) and spin ($s$) channels can be detected via the momentum-dependent static structure factor (see, e.g., Refs.~\cite{punk2014topological, PhysRevB.97.014423, PhysRevB.104.165127})
\begin{align}
S^{c/s}(\qv) = \sum_{ll'} e^{-i\qv \cdot ({\rm\bf R}_l - {\rm\bf R}_{l'})} X^{c/s}_{ll'}(\qv, \omega=0)\,,
\label{eq:structure}
\end{align}
where the vector ${\rm\bf R}_{l}$ depicts the position of the atom $l$ within the unit cell. 
In the high-temperature ${\sqrt{3}\times\sqrt{3}}$ phase, where ${l=l'}$, the static structure factor coincides with the static susceptibility ${X^{c/s}(\qv, \omega=0)}$ obtained at zero frequency ${\omega}$. 
The divergence of the structure factor at momenta ${\bf q= Q}$ indicates a transition to a symmetry-broken ordered state associated with Bragg peaks at ${\bf Q}$.
Transitions without symmetry-breaking, such as the metal to Mott insulator phase transition, can be observed by inspecting the spectral function.
In this work, the introduced many-body problem~\eqref{eq_hamilt} is solved using the dual triply irreducible
local expansion (\mbox{D-TRILEX}) method~\cite{PhysRevB.100.205115, PhysRevB.103.245123, 10.21468/SciPostPhys.13.2.036}. 
This method provides a consistent treatment of the local correlation effects and the non-local collective electronic fluctuations in the charge and spin channels~\cite{PhysRevLett.127.207205, stepanov2021coexisting, PhysRevLett.129.096404, 2022arXiv220402116V}.
Importantly, \mbox{D-TRILEX} is also able to account for the long-range Coulomb interaction~\cite{stepanov2021coexisting} and the SOC~\cite{10.21468/SciPostPhys.13.2.036}, which are the two important aspects of the considered material. 
More details on the many-body calculations are provided in the Methods section.

\paragraph{\bf Phase diagram for the ${\sqrt{3} \times \sqrt{3}}$ structure.}

\begin{figure}[b!]
    \includegraphics[width=\linewidth]{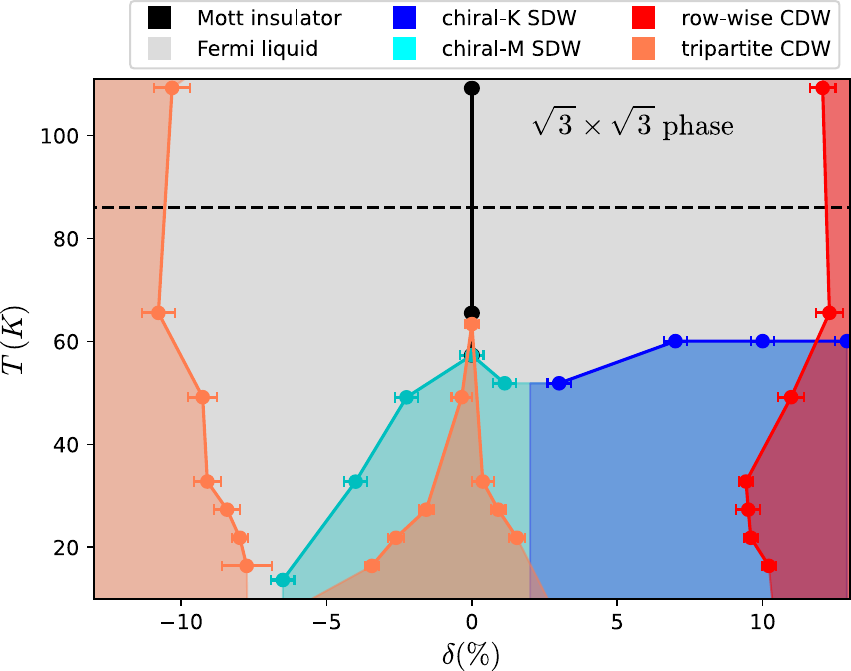} 
    \caption{Phase diagram for Si(111):Pb in the ${\sqrt{3}\times\sqrt{3}}$ structure. Different phases as a function of doping $\delta$ and temperature $T$ are highlighted by colors (color code in the legend). Calculations have been performed by fixing the temperature and conducting a scan over doping levels on a finite grid, which defines the error bars. Positive (negative) values of $\delta$ correspond to electron (hole) doping. The horizontal dashed black line depicts the temperature ${T=86~\text{K}}$ at which the material exhibits a structural phase transition according to Refs.~\cite{PhysRevLett.94.046101, PhysRevB.75.155411}. The vertical line that divides the magnetic phases below the transition points is only meant as a guide to the eye, since we are not able to distinguish between the two different phases in symmetry-broken regime.}
    \label{fig:phase_diagram}
\end{figure}

The phase diagram for the Si(111):Pb material in the ${\sqrt{3}\times \sqrt{3}}$ structure is shown in Fig.~\ref{fig:phase_diagram} as a function of doping level $\delta$ and temperature $T$. In the considered system, the value of the local Coulomb interaction is approximately $3$ times larger than the electronic bandwidth~\cite{PhysRevLett.110.166401, PhysRevB.94.224418}. 
As a consequence, at high temperature the half-filled system lies deep in the Mott insulating phase (black line at ${\delta=0\%}$). 
A small amount of hole- or electron-doping causes a phase transition to a Fermi liquid regime (gray area). 
For this reason, the electronic behavior in doped Si(111):Pb is a characteristic manifestation of the physics of a doped Mott insulator.

Upon solving the many-body problem~\eqref{eq_hamilt} we identify several different spin density wave (SDW) and CDW orderings at different values of doping, as illustrated in Fig.~\ref{fig:phase_diagram}. 
Since these phases are realized for a non-integer filling of electrons, they are likely metallic. 
However, we cannot confirm this in our actual calculations because our method does not allow us to perform calculations inside phases induced by dynamic symmetry breaking.
Specifically, around half-filling we observe a CDW ordering (orange area around ${\delta=0\%}$) characterised by the divergence of the static charge structure factor at the ${{\bf Q}={\rm K}}$ point of the Brillouin zone (BZ).
This ordering is analogous to the 120$^\circ$-N\'eel phase of the Heisenberg model on a triangular lattice with three inequivalent sites in the unit cell (see, e.g., Ref.~\onlinecite{PhysRevLett.86.1106}).
For this reason, hereinafter we call this type of ordering a ``tripartite CDW''. Importantly, we find that this instability does not appear if instead of the full long-range Coulomb potential $V_{ij}$ one considers the interaction only between nearest-neighbour lattice sites.
In the presence of only local interactions, the Mott phase and a CDW would be mutually exclusive. Here, we note that the effective long-range interaction is enhanced by correlations as the temperature is reduced, while the local interaction barely depends on temperature.
We would also like to note that competing tripartite CDW and Mott phases have been observed experimentally in the other adatom system Ge(111):Sn~\cite{PhysRevB.88.125113}.

\begin{figure}[t!]
    \centering
    \includegraphics[width=\linewidth]{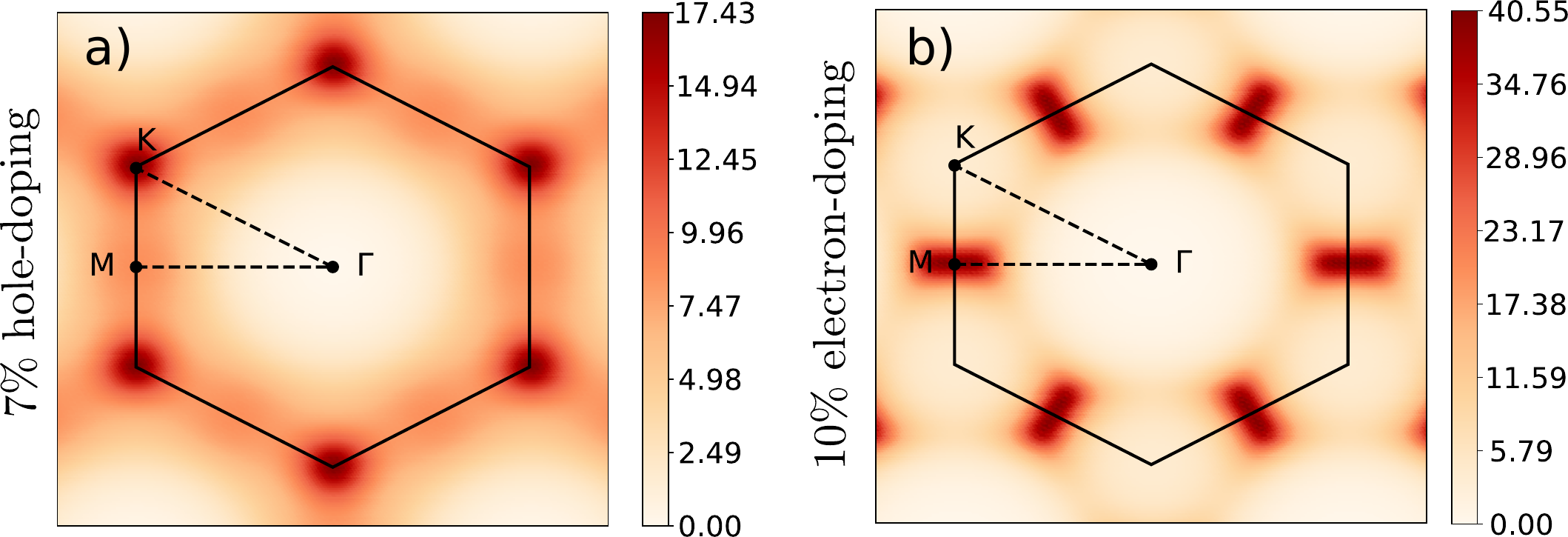}
    \caption{The static charge structure factor $S^{c}({\bf q})$. The result is obtained close to the CDW transition points ${\delta=-7\%}$, ${T=25~{\rm K}}$ (a) and ${\delta=10\%}$, ${T=67~{\rm K}}$ (b). In the hole-doped case, the Bragg peaks in the structure factor appear at the ${{\bf Q}={\rm K}}$ points of the BZ indicating the tripartite CDW ordering. In the electron-doped case, the ordering vector ${{\bf Q}={\rm M}}$ corresponds to the row-wise CDW instability.}   \label{fig:chi_charge}
\end{figure}

Additionally, we identify two other CDW phase transitions at dopings around ${\delta=\pm10\%}$.
These instabilities appear to be weakly temperature-dependent and approximately symmetric with respect to half-filling. 
At hole doping, the CDW ordering vector remains ${{\bf Q}={\rm K}}$ (orange area), as in the half-filled case.
However, in the electron-doped regime the divergence of the static charge structure factor occurs at the ${{\bf Q}={\rm M}}$ point of the BZ, which can be associated with a ``row-wise CDW'' ordering (red area). 
One can speculate, that this ordering might be related to the isoelectronic mosaic phase observed in Si(111):Pb~\cite{PhysRevB.66.035325} or to the intermediate stripe-like order in the alkali doped Si(111):Sn surface~\cite{PhysRevLett.124.097602}.
However, a direct observation of the row-wise CDW phase in Si(111):Pb has not been performed yet.
The momentum-resolved static charge structure factors obtained close to both these CDW instabilities are shown in Fig.~\ref{fig:chi_charge}, where the Bragg peaks clearly indicate the corresponding ordering vectors.

\begin{figure}[t!]
    \centering
    \includegraphics[width=\linewidth]{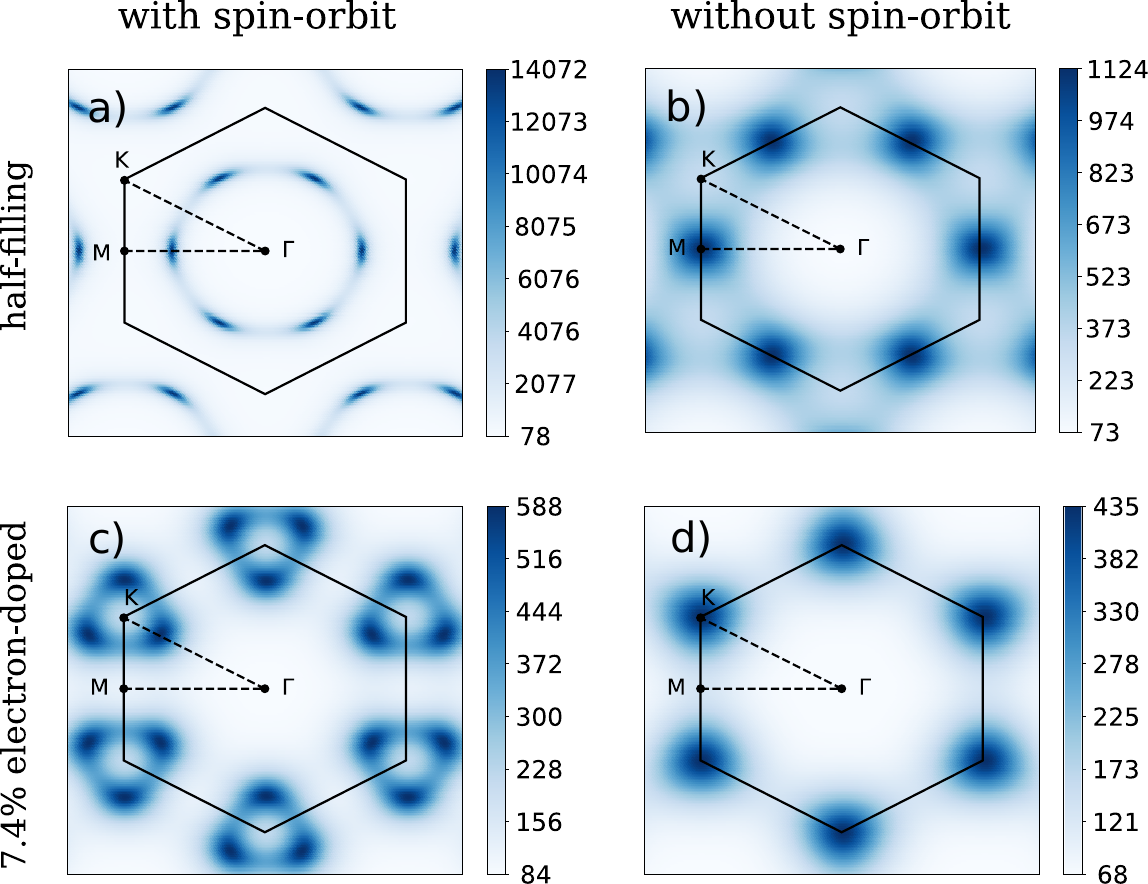}
   \caption{The static spin structure factor $S^{s}({\bf q})$. The results are obtained with (left column) and without (right column) SOC, respectively. The upper row corresponds to the half-filling ${\delta=0\%}$, the bottom row to $\delta=7.4\%$ electron doping. The chosen temperature, ${T=67~\text{K}}$, is close to the SDW transition. Without SOC, the Bragg peaks in $S^{s}({\bf q})$ indicate the row-wise (b) and the N\'eel (d) magnetic structures. Taking into account the SOC, the Bragg peaks in both cases shift to incommensurate positions with \mbox{chiral-M} (a) and \mbox{chiral-K} (c) SDW orderings.
   }
    \label{fig:SO_comp_SP}
\end{figure}

In addition to the CDW instabilities, we also observe magnetic structures with different ordering vectors depending on the doping level (cyan and blue areas in Fig.~\ref{fig:phase_diagram}). 
Around half-filling, we observe a SDW characterized by Bragg peaks in the static spin structure factor that lie at an incommensurate point ${{\bf Q} \simeq \frac{2}{3} \,{\rm M}}$ 
of the BZ (Fig.~\ref{fig:SO_comp_SP}\,a). 
At ${\delta\gtrsim2\%}$ of electron-doping the SDW ordering vector changes, and the peaks shift to another incommensurate position ${{\bf Q} \simeq \frac{3}{4} \,{\rm K}}$
(Fig.~\ref{fig:SO_comp_SP}\,c).
The appearance of the Bragg peaks at incommensurate points of the BZ signals the formation of a chiral magnetic order  that can be viewed as a superposition of spin spirals. 
According to the position of the Bragg peaks, we call these magnetic structures ``\mbox{chiral-M}'' (cyan area) and ``\mbox{chiral-K}'' (blue area) SDW, respectively. 
The presence of the chiral magnetic orderings in Si(111):Pb suggests that this material might be a suitable candidate for the realization of skyrmionic phases that can possibly be stabilized under an external magnetic field~\cite{PhysRevB.94.224418}.

Remarkably, the obtained chiral SDW structures partially coexist with the CDW orderings.
In the considered Si(111):Pb material such coexistence was recently observed by means of STM measurements~\cite{PhysRevLett.120.196402}, but an estimate of the doping level in the system was not provided, presumably due to difficulties in the determination of the effective doping. 
Remarkably, we find that the \mbox{chiral-M} SDW structure coexists only with the tripartite CDW ordering, which appears around half-filling.
Instead, the row-wise CDW ordering coexists only with the \mbox{chiral-K} SDW at a relatively large electron doping.
This observation suggests a simple way for a qualitative estimation of the doping level in the experimentally measured material, which is difficult to probe directly (see Refs.~\cite{PhysRevLett.119.266802, PhysRevLett.125.117001} and related supplemental materials for discussion). 

We have made a very crude estimation of the doping level by calculating the area of the Fermi surface that can be deduced from the STM map shown in Ref.~\cite{PhysRevLett.120.196402}.
The obtained result is compatible with up to $\simeq11\%$ electron-doping, which coincides with the region of coexisting chiral-K SDW and row-wise CDW orderings. This result appears to be consistent with use of an electron-doped substrate~\cite{PhysRevLett.120.196402_sm}.

\paragraph{\bf Effect of the SOC.}

We observe, that the large SOC, which is an intrinsic feature of Si(111):Pb, manifests itself in the magnetic properties of the material. 
In particular, the effect of the SOC can be seen in the spin structure factors shown in Fig.~\ref{fig:SO_comp_SP}. As we have shown above, the SOC results in the formation of the \mbox{chiral-M} (a) and \mbox{chiral-K} (c) SDW orderings in the system.
Instead, if the SOC is not taken into account, the Bragg peaks in the static spin structure factor calculated close to the SDW phase transitions appear at the ${{\bf Q}=\text{M}}$ (b) and ${{\bf Q}=\text{K}}$ (d) points of the BZ.
These instabilities correspond to commensurate row-wise and N\'eel magnetic structures, respectively.  
Remarkably, despite the shift of the peaks in the BZ and the consequent change of the ordering of the system, we find that the position of the phase boundaries is not affected by the SOC (up to the error bars of our calculations), similarly to what has been found in Ref.~\onlinecite{2022arXiv221009384B} for a square lattice. 
Based on this result, one can argue that the phase boundaries in the considered system can be obtained correctly without taking into account the SOC. 
However, considering the SOC is absolutely necessary for an accurate determination of the ordering vectors.

\paragraph{\bf Effective Heisenberg model.}

The observed changes in the spin structure factor as a function of doping level can be explained by analyzing the exchange interactions~\cite{PhysRevLett.121.037204, PhysRevB.99.115124, PhysRevB.105.155151}.
These quantities are accessible in \mbox{D-TRILEX} calculations~\cite{10.21468/SciPostPhys.13.2.036}.
To this aim, we consider the following effective Heisenberg-like classical spin Hamiltonian with bilinear magnetic exchange interactions:
\begin{align}
H = J\sum_{\langle ij \rangle}{\bf S}_i \cdot {\bf S}_j + J'\sum_{\langle \langle ij \rangle \rangle}{\bf S}_i \cdot {\bf S}_j + D\sum_{\langle ij\rangle} {\bf z} \cdot \left({\bf S}_i \times {\bf S}_j\right)\,.
\label{eq:Hspin}
\end{align}
In this expression, $J$ and $J'$ are the nearest-neighbor ${\langle ij \rangle}$ and the next-nearest-neighbor ${\langle\langle ij \rangle\rangle}$ exchange interactions, respectively.
$D$ is the nearest-neighbor Dzyaloshinskii–Moriya interaction (DMI), which appears due to the SOC. We have also calculated the symmetric anisotropy, but we omit it for simplicity as it hardly affects the following considerations. The value of its only non-zero component is ${\Gamma_{yy} \approx 0.5D}$ in the whole range of $\delta$ considered here.    

\begin{figure}[t!]
    \centering
    \includegraphics[width=\linewidth]{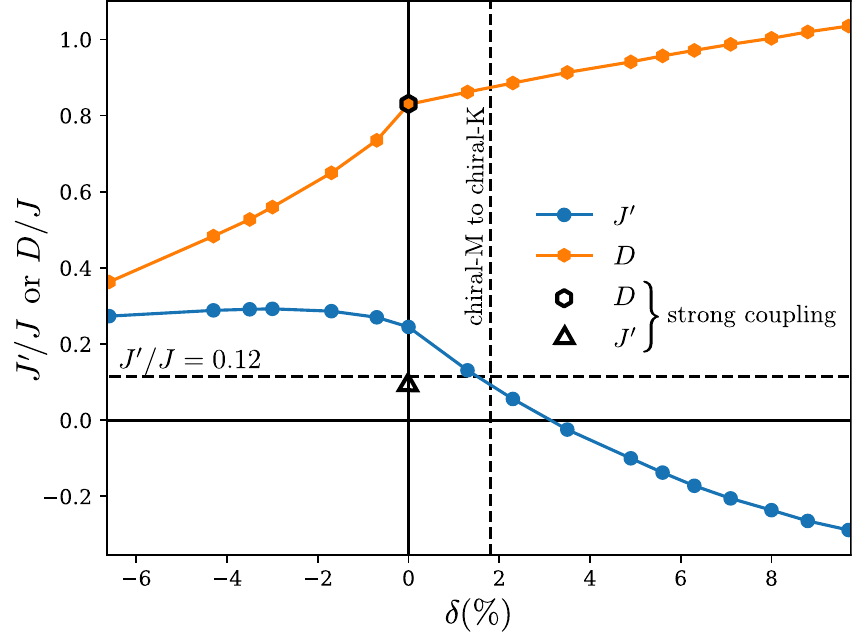}
     \caption{Magnetic exchange interactions as a function of doping. The orange line depicts the value of the nearest-neighbor Dzyaloshinskii–Moriya interaction $D/J$. The blue line corresponds to the next-nearest-neighbor exchange interaction $J'/J$. Both quantities are normalized by the value of the nearest-neighbor exchange $J$. The results are obtained at ${T=50 \; {\rm K}}$. The black hexagon and the black dot represent the values of $D/J$ and $J'/J$ obtained at half-filling in Ref.~\cite{PhysRevB.94.224418} using the strong-coupling approximation. The vertical dashed black line at ${\delta=1.8 \%}$ indicates the transition from the \mbox{chiral-M} to the \mbox{chiral-K} phases according to our calculations. The horizontal dashed line at ${J'/J=0.12}$ represents the prediction for the M to K transition in the $J$-$J'$ Heisenberg model obtained from Monte Carlo calculations in Ref.~\onlinecite{Ramazanov2011}.}
    \label{fig:exchange_int}
\end{figure}

Fig.~\ref{fig:exchange_int} shows the evolution of $J'$ and $D$, normalized by the value of $J$, as a function of doping.
Remarkably, we find that the magnitude of $D$ in Si(111):Pb is of the order of the nearest-neighbor exchange interaction $J$, which is very unusual for magnetic systems.
Moreover, $D$ and $J$ even become equal in the electron-doped case.
At half-filling the value of $D/J$ coincides with the one obtained in Ref.~\cite{PhysRevB.94.224418} using the strong-coupling approximation.
This fact confirms that the half-filled Si(111):Pb material lies in the strong-coupling regime.
Further, we observe that the ratio $D/J$ has an approximately linear dependence on doping with different slopes in the hole- and electron-doped regimes. 
In the hole-doped case, $D/J$ substantially decreases upon increasing the doping.
Instead, in the electron-doped regime, $D/J$ slowly increases with increasing $\delta$. 
This behavior explains the formation of the chiral SDW orderings in the regime of doping levels ${\delta\gtrsim-7\%}$, where DMI is strong enough (${D/J\gtrsim0.4}$) to be able to shift the Bragg peaks from a commensurate to an incommensurate position, as shown in Fig.~\ref{fig:SO_comp_SP}.

While DMI is responsible for the formation of chiral spin structures, the change in the ratio $J'/J$ with doping explains the transformation of the magnetic ordering from the M- to the K-type, as observed in our calculations. 
The magnitude of $J'$ is rather small compared to $J$ and $D$, but it is not negligible.
In addition, we find that the actual value of the more distant, next-nearest-neighbor exchange interaction $J'$ is substantially larger than the one predicted by a strong-coupling estimate~\cite{PhysRevB.94.224418}.
An important feature is that the ratio $J'/J$ is nearly constant in the hole-doped regime,
while in the electron-doped case it substantially decreases and even changes sign.
We attribute this variation of $J'/J$ to the shift of the Bragg peaks in the spin structural factor from M to K, which is consistent with Monte Carlo calculations for the $J$-$J'$ Heisenberg model on a triangular lattice performed in Ref.~\cite{Ramazanov2011}.
It has been shown there, that the transition from a row-wise (${{\bf Q}=\text{M}}$) to a N\'eel (${{\bf Q}=\text{K}}$) magnetic order occurs for ${J'/J \simeq 0.12}$.
As shown in Fig.~\ref{fig:exchange_int}, this result coincides with our estimate for the transition point between the \mbox{chiral-M} to \mbox{chiral-K} SDW orderings.
In this figure, the horizontal dashed black line depicts the ${J'/J = 0.12}$ value, and the vertical dashed black line marks the mean-point between the closest doing levels that correspond to \mbox{chiral-M} and \mbox{chiral-K} SDW orderings.

\begin{figure}[t!]
\centering
\includegraphics[width=\linewidth]{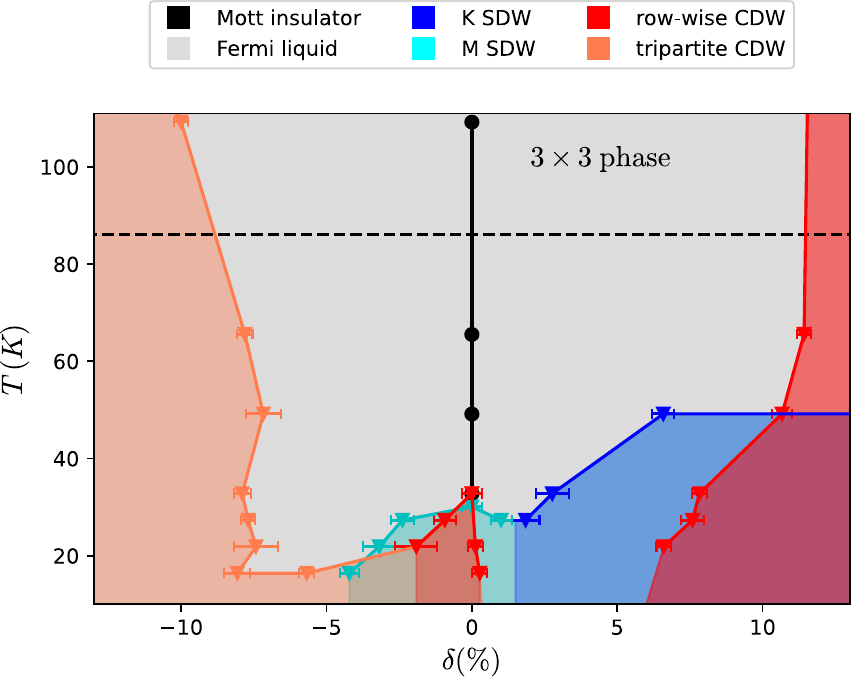}
\caption{Phase diagram for Si(111):Pb in the ${3\times3}$ reconstruction. Different phases that appear in the system as a function of the doping $\delta$ and temperature $T$ are highlighted in color. The color code can be found in the legend. Calculations have been performed by fixing $T$ and conducting a scan over doping levels on a finite grid, which defines the error bars. The vertical line dividing the magnetic phases below the transition points are only meant as a guide to the eye. 
} 
\label{fig:phases_3x3}
\end{figure}

\paragraph{\bf Phase diagram for the ${3 \times 3}$ reconstruction.}

At low temperature, Si(111):Pb undergoes a structural phase transition from ${\sqrt{3}\times \sqrt{3}}$ to ${3\times 3}$ periodicity. 
The ${3\times 3}$ reconstruction exhibits a \mbox{1-up-2-down} configuration of Pb adatoms, as confirmed in experiments~\cite{ PhysRevLett.94.046101, PhysRevB.75.155411} and by DFT calculations~\cite{PhysRevLett.98.086401, PhysRevLett.120.196402}.
In order to account for the effect of the structural phase transition, we also perform many-body calculations for the ${3\times 3}$ reconstruction of adatoms. 
The \mbox{1-up-2-down} configuration requires to consider a unit cell with three Pb atoms, which significantly increases the cost of the numerical calculations.
As previously discussed, the inclusion of the SOC does not affect the position of the phase boundaries in the considered material. 
In order to make numerical calculations in the ${3\times 3}$ phase feasible, we neglect the Rashba term in the model Hamiltonian~\eqref{eq_hamilt}.

\begin{figure}[t!]
    \centering
  \includegraphics[width=\linewidth]{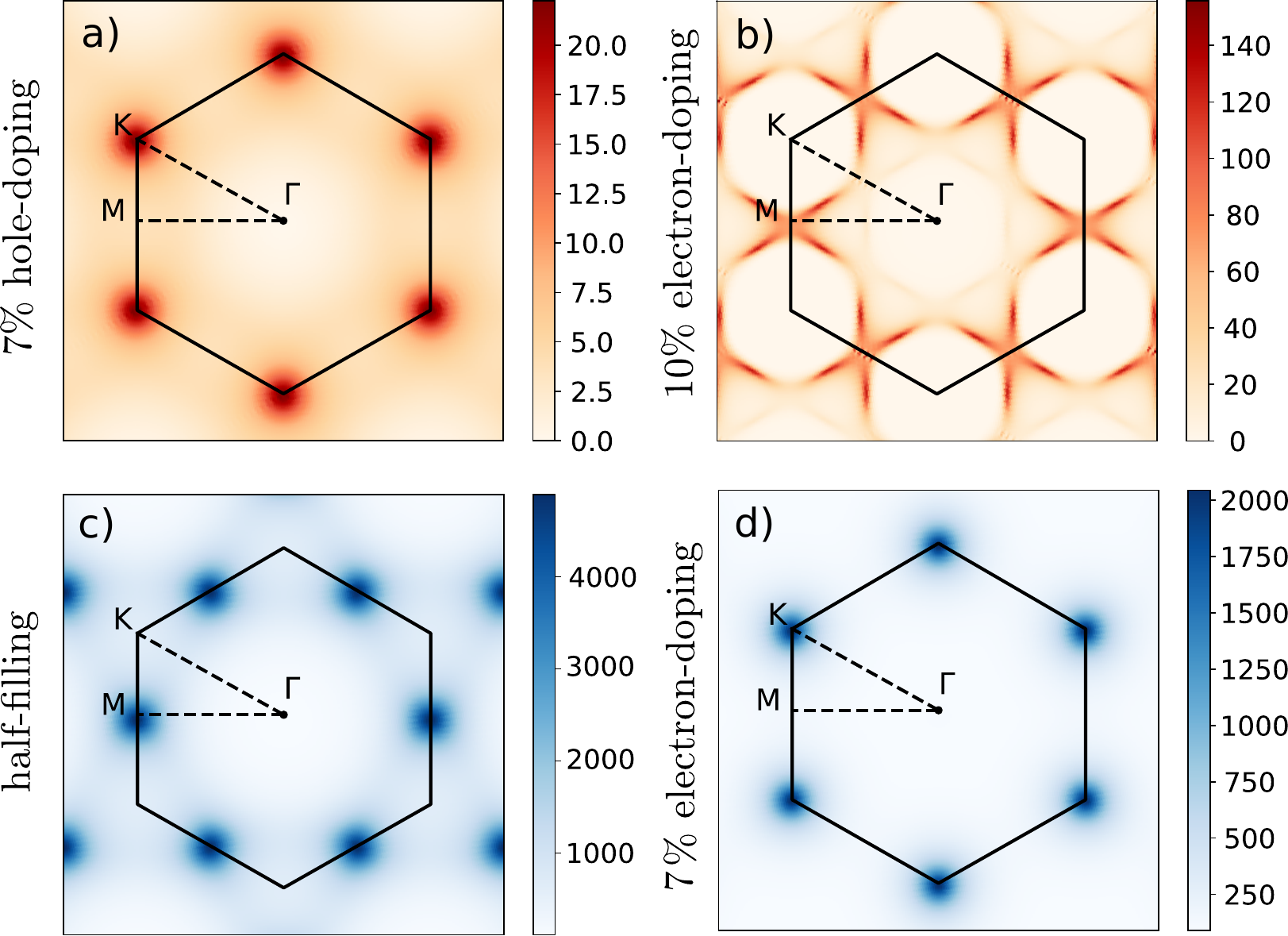}
    \caption{Charge (top row) and spin (bottom row) static structure factors. The depicted results are for the ${3\times3}$ reconstruction in the vicinity of the tripartite CDW (a), row-wise CDW (b), M SDW (c), and K SDW (d) phase transitions. The corresponding doping levels are specified for each panel.
    The temperature is chosen to be close to the phase boundaries,  $T \approx 25 \, {\rm K}$ in panel (a), $T\approx 35 \, {\rm K}$ in panel (c) and $T\approx 67 \, {\rm K}$ in (b) and (d).
   }
    \label{fig:structure_factors}
\end{figure}

Fig.~\ref{fig:phases_3x3} shows the resulting phase diagram for the ${3\times 3}$ reconstruction, which qualitatively agrees with the one obtained for the ${\sqrt{3}\times \sqrt{3}}$ structure.
Indeed, the phase diagram for the \mbox{1-up-2-down} configuration of Pb atoms also contains row-wise and tripartite CDW phases that are nearly temperature-independent and appear at values of the hole- and electron-doping comparable to the ${\sqrt{3}\times \sqrt{3}}$ case.
We note that these dynamical CDW instabilities emerge on top of the structural phase transition, which affects the ordering vector of the row-wise CDW structure.
Indeed, Fig.~\ref{fig:structure_factors}\,b shows that the Bragg peaks in the charge structure factor are now found at incommensurate positions in the vicinity of the M point of the BZ.
This result can be explained by the observation that the divergence of the corresponding charge susceptibility ${X^{c}_{ll'}({\bf q}, \omega=0)}$, which enters the expression~\eqref{eq:structure} for the structure factor, also appears at incommensurate positions in the vicinity of the M point of the reduced BZ.
A wave-vector at the M point would mean row-wise ordering, as in the single-site case. However, here we have two overlapping orderings: a row-wise order induced by correlations and the $3\times 3$-underlying broken symmetry due to the lattice distortion. The reason for this pattern is that a perfect row-wise arrangement would not be commensurate with the underlying \mbox{1-up–2-down} structure.
It means that the spontaneous symmetry breaking leading to the row-wise CDW ordering occurs between different unit cells on the lattice, but not within the unit cell of three Pb atoms.
On the contrary, we find that the ordering vector ${{\bf Q}=\text{K}}$ of the tripartite CDW instability remains unchanged upon the structural transition (top left panel of Fig.~\ref{fig:structure_factors}). 
The tripartite CDW corresponds to the ordering, where all three Pb atoms in the unit cell are inequivalent.
The fact that upon the tripartite CDW phase transition the charge susceptibility diverges at the $\Gamma$ point of the reducible BZ confirms the statement that in this case the spontaneous symmetry breaking occurs within the unit cell.
Consequently, the \mbox{1-up-2-down} structure of Pb atoms in the unit cell transforms to a tripartite structure, and the Bragg peaks in the structural factor appear at the K point of the BZ as usual. 

The structural transition also affects the phase boundaries of the temperature-dependent instabilities.
All of them, namely the CDW around half-filling and both SDW instabilities, are pushed down to lower temperatures. 
This can be related to the appearance of an effective local potential $\Delta_{l}$ upon the structural transition to the \mbox{1-up-2-down} structure.
This potential acts as an on-site doping that differs from site to site and thus suppresses collective charge and spin fluctuations.
Interestingly, the CDW ordering found around half-filling in the ${3\times3}$ reconstruction has a row-wise structure instead of the tripartite one observed in the ${\sqrt{3}\times\sqrt{3}}$ case. 
As discussed above, the row-wise ordering does not break the \mbox{1-up-2-down} structure of Pb adatoms in the unit cell.
Probably for this reason the formation of the row-wise CDW is more favorable in the ${3\times3}$ phase.
Finally, we note that apart of decreasing the critical temperature for the SDW instabilities, the structural transition does not affect the magnetic ordering in the system. 
As in the ${\sqrt{3}\times\sqrt{3}}$ case we find the M SDW ordering around half-filling and the K SDW ordering at ${\delta\gtrsim2\%}$ of electron-doping.
In our calculations, the Bragg peaks in the corresponding spin structure factors appear at commensurate ${{\bf Q}=\text{M}}$ (top left panel of Fig.~\ref{fig:structure_factors}) and ${{\bf Q}=\text{K}}$ (top right panel of Fig.~\ref{fig:structure_factors}) positions.
We expect that the inclusion of the SOC would shift the peaks to incommensurate positions and lead to the formation of the chiral magnetic structures also in the ${3\times3}$ case.

\section{Discussion}

We performed many-body calculations for a system of Pb adatoms on a Si(111) substrate, including the SOC and long-range Coulomb interactions. 
By investigating spatial collective electronic fluctuations in both, charge and spin channels, we observe a rich variety of different symmetry-broken charge- and spin-density wave phases in the low temperature regime by varying the doping level. Regarding the Mott physics, our results show a picture similar to that of Sn on Si(111): the system is a Mott insulator at half-filling, but immediately turns into a metal as soon as some small doping is introduced in the system~\cite{PhysRevLett.125.117001}.
We find that the strong SOC in this material results in a very large Dzyaloshinskii–Moriya interaction comparable to the usual Heisenberg exchange interaction.
This leads to the formation of \mbox{chiral-M} and \mbox{chiral-K} SDW phases, a signature of which have recently been observed in STM measurements~\cite{PhysRevLett.120.196402}. 
These chiral spin structures are compatible with magnetic skyrmion textures, as highlighted in previous theoretical calculations~\cite{PhysRevB.94.224418}. 
Tuning the doping level allows one to switch between the two chiral SDW phases and thus realize different kinds of spin structures with potential topological structure in one material. 
We note that a similar change of the magnetic ordering was proposed for a Si(111):Sn system by means of varying the local Coulomb interaction~\cite{PhysRevB.83.041104}.

We also find that two different CDW orderings can appear in Si(111):Pb, and that their geometry is strongly affected by the doping level. 
The values of doping, at which the transition takes place, appear to be consistent with the intrinsic doping levels observed in this kind of systems~\cite{PhysRevLett.119.266802}. 
There is an on-going debate whether the ${3 \times 3}$ pattern of charge densities observed in experiments emerges in Si(111):Pb due to a dynamical symmetry breaking associated with strong electronic correlations~\cite{PhysRevLett.123.086401}, or by means of a structural transition~\cite{PhysRevB.104.045126}.
We argue that the corresponding 1-up-2-down structure of Pb adatoms can be realized in the system upon either the structural transition from the ${\sqrt{3}\times\sqrt{3}}$ to the ${3\times3}$ phase, or the dynamical symmetry breaking towards the row-wise CDW phase, depending on the doping level and temperature.
In addition, we find another CDW ordering in the system associated with the formation of a tripartite structure.

In order to realize these theoretically predicted phases in experiment, it is necessary to use a probe sensitive to collective excitations, as well as to be able to give an accurate estimation of the doping level. 
Since the precise occupation of the isolated band is experimentally challenging to access, we propose an alternative way to identify the doping level. 
Using a probe sensitive to the underlying magnetic structure, such as spin-polarized STM~\cite{Serrate2010}, could prove a valid alternative to the measurements of the doping, since the magnetic textures appearing at different doping levels exhibit different geometry and also coexist with different types of CDW ordering. 

A recent study on a similar adatom system of Sn adatoms on germanium indicated the presence of strong electron-phonon coupling (EPC)~\cite{PhysRevB.107.045303}. This system has a different composition, so it is not known if a similar effect holds also for Pb on Si(111). We argue that the EPC scales as $1/\sqrt{M}$ with the atomic mass $M$, so the contribution to the effective electron-electron interaction scales as $1/M$ and it is much smaller on the Pb surface  than in the case of Sn. Additionally, in order to strongly affect the properties of the system, EPC would need to overcome the very strong Coulomb interaction present in this system. As this is very unlikely to occur, we conclude that we do not expect this contribution to be crucial to determine the phases of this system. However, it could modify the position of the phase boundaries, hence in the future it would be desirable to devise a way to deal with EPC in D-TRILEX calculations. Further studies are also required in order to investigate superconductivity in the low-temperature regime.

\section{Method}
\label{sec:Methods}

\paragraph{\bf Ab-initio DFT calculations.}

All model parameters used in the model Hamiltonian~\eqref{eq_hamilt} have been obtained from \textit{ab-initio} calculations. 
For the ${\sqrt{3} \times \sqrt{3}}$ structure of adatoms we adapted the parameters from Ref.~\onlinecite{PhysRevB.94.224418}, where a Wannier projection on localized orbitals was performed to obtain the nearest-neighbor ${t_{01} = 41.3~\text{meV}}$ and the next-nearest-neighbor ${t_{02} = -19.2~\text{meV}}$ hopping amplitudes. 
The Rashba parameters ${\gamma_{01} = 16.7~\text{meV}}$ and ${\gamma_{02} = 2.1~\text{meV}}$ are taken from the same work as the hopping amplitudes. 
The value of the local Coulomb interaction ${U = 0.9~\text{eV}}$ is the one obtained from cRPA calculations~\cite{Hansmann_2013, PhysRevLett.110.166401, PhysRevB.94.224418}. 
The long-range Coulomb interaction with a realistic $1/r$ tail is parametrized by the nearest-neighbor interaction ${V_{01}=0.5~\text{eV}}$ as suggested in Refs.~\onlinecite{Hansmann_2013, PhysRevLett.110.166401, PhysRevLett.101.126401}. 
The direct exchange interaction between neighboring sites that enters Eq.~\eqref{eq_hamilt} is rather small and reads ${J_{\langle ij \rangle}=1.67~\text{meV}}$~\cite{PhysRevB.94.224418}. 

For the ${3\times 3}$ reconstruction, we simulated the surface by a slab geometry consisting of $1/3$ monolayer of Pb adatoms on top of three Si bi-layers, as established in previous works~\cite{Profeta2007, PhysRevB.82.035116, Hansmann_2013, PhysRevB.94.224418, PhysRevLett.120.196402}. 
The Pb adatoms occupy the T$_4$ positions. 
The dangling bonds of the bottom Si bi-layer are compensated by hydrogen capping, and \unit[19]{\AA} of vacuum are included in the simulation. 
For structural relaxations we employ the WIEN2k~\cite{wien2k,wien2k2020} program package, a full-potential linearized-augmented plane-wave code.  
We start with the relaxation of the $\sqrt{3}\times\sqrt{3}$ structure, which contains one Pb per unit cell. 
We then construct the ${3\times3}$ supercell containing 3 Pb atoms (66 atoms in total, thereof 54 Si). 
To relax the ${3\times3}$ structure, which in experiment is found in a \mbox{1-up-2-down} configuration, we displace one of the three Pb adatoms by \unit[0.4]{\AA} perpendicularly to the surface in the first DFT self-consistent-field iteration.
We then let the internal coordinates of all atoms in the supercell relax freely until convergence. 
We employed a  multisecant approach~\cite{wien2k2020}, as implemented in WIEN2k~\cite{wien2k,wien2k2020}. 
A k-grid with $6\times6\times1$ k-points in the reducible Brillouin zone was used and internal coordinates were relaxed until forces were less than \unit[2]{mRy/bohr}. 
We employed the generalized gradient approximation (PBE), spin-orbit coupling was neglected. 
In agreement with experiment, we find the stabilisation of a ${3\times3}$ reconstruction, where one Pb adatom is vertically displaced by \unit[0.22]{\AA} compared to the other two Pb adatoms in the supercell. 
The energy gain of this \mbox{1-up-2-down} reconstruction is found to be \unit[9.5]{meV} with respect to a flat adatom layer. 
These findings are in good agreement with previous \textit{ab initio} calculations~\cite{PhysRevLett.94.046101,PhysRevLett.120.196402}.
We find that the computed band structure for the ${3\times3}$ reconstruction can be well interpolated with a 3-band dispersion using the same parameters taken from Ref.~\onlinecite{PhysRevB.94.224418} by simply adding a local potential $\Delta_l$ to the inequivalent Pb atoms ${l\in\{1,2,3\}}$ in the model Hamiltonian. We choose this approach to ensure better comparability between the calculations.
The obtained values for the potential are ${\Delta_1 = \Delta_2 = 31.5~\text{meV}}$ and ${\Delta_3=-55.4~\text{meV}}$. The effect of the substrate-induced deformation, which corresponds to a static electron-phonon interaction, can be a crucial ingredient for the formation of the 3x3 structure~\cite{PhysRevB.104.045126}.
We stress that this effect of phonons is taken into account in our calculations of the 3x3 structure by keeping the lattice distortion appearing at the DFT level in the interacting problem~\eqref{eq_hamilt}.

\paragraph{\bf Many-body D-TRILEX calculations.}

The interacting electronic problem~\eqref{eq_hamilt} is solved using the finite temperature \mbox{D-TRILEX} method~\cite{PhysRevB.100.205115, PhysRevB.103.245123, 10.21468/SciPostPhys.13.2.036}.
To this aim, we first perform converged dynamical mean-field theory (DMFT) calculations~\cite{RevModPhys.68.13} with the w2dynamics package~\cite{WALLERBERGER2019388} in order to take into account local correlation effects in a numerically exact way.
Furthermore, the effect of the non-local collective electronic fluctuations and of the SOC is taken into account diagrammatically as described in Ref.~\onlinecite{10.21468/SciPostPhys.13.2.036}.
The spin susceptibility ${X^{s}_{ll'}(\qv,\omega)}$ required for the calculation of the structure factor~\eqref{eq:structure} is defined as the maximum eigenvalue of the matrix
\begin{align}
X^{ss'}_{ll'}(\qv,\omega) = \langle S^s_{l\qv\omega} S^{s'}_{l',-\qv, -\omega}\rangle
\end{align}
in the space of spin channel indices ${s^{(\prime)} \in \{s_x, s_y, s_z\}}$.
The charge susceptibility is defined as:
\begin{align}
X^{c}_{ll'}(\qv,\omega) = \langle n_{l\qv\omega} n_{l'-\qv, -\omega}\rangle\,.
\end{align}
Note that in this work the susceptibility is computed non-self-consistently, as, e.g., in Ref.~\onlinecite{stepanov2021coexisting}.
This means that the susceptibility is calculated on the basis of the electronic Green's functions dressed only by the local DMFT self-energy, which resembles the way the susceptibility is computed in DMFT~\cite{PhysRevB.103.245123}.
This procedure allows one to treat collective electronic instabilities in the charge and spin channels independently without mutually affecting each other. 

The magnetic exchange interactions used to construct the effective Heisenberg model~\eqref{eq:Hspin} are also computed within the \mbox{D-TRILEX} scheme as explained in Ref.~\cite{10.21468/SciPostPhys.13.2.036}. \\

{\bf Competing Interests.} The Authors declare no Competing Financial or Non-Financial Interests.

{\bf Author contributions.} All authors discussed the results and contributed to the preparation of the manuscript.
 
{\bf Funding.} M.V., A.R. and A.I.L. acknowledge the support by the Cluster of Excellence ``Advanced Imaging of Matter'' of the Deutsche Forschungsgemeinschaft (DFG) - EXC 2056 - Project No.~ID390715994 and -SFB-925 - Project No.~170620586.
M.V., E.A.S., and A.I.L. also acknowledge the support by North-German Supercomputing Alliance (HLRN) under the Project No.~hhp00042.
A.R. acknowledges support by
the European Research Council (ERC-2015-AdG-694097), Grupos Consolidados (IT1249-19)
and the Flatiron Institute, a division of the Simons Foundation.
S.B. acknowledges the support from IDRIS/GENCI Orsay under project number A0130901393.
The work of E.A.S. was supported by the European Union’s Horizon 2020 Research and Innovation programme under the Marie Sk\l{}odowska Curie grant agreement No.~839551 - $\text{2DMAGICS}$.

{\bf Data Availability.} The data that support the findings of this work are available from the corresponding author upon reasonable request.

{\bf Code Availability.} First-principle calculations have been performed on the basis of the WIEN2k~\cite{wien2k,wien2k2020} program package that can be requested from the developers. 
Many-body calculations have been performed using the implementation of the \mbox{D-TRILEX} method~\cite{10.21468/SciPostPhys.13.2.036} that can be obtained from the corresponding author upon reasonable request.

\bibliography{main}

\begin{thebibliography}{80}%
\makeatletter
\providecommand \@ifxundefined [1]{%
 \@ifx{#1\undefined}
}%
\providecommand \@ifnum [1]{%
 \ifnum #1\expandafter \@firstoftwo
 \else \expandafter \@secondoftwo
 \fi
}%
\providecommand \@ifx [1]{%
 \ifx #1\expandafter \@firstoftwo
 \else \expandafter \@secondoftwo
 \fi
}%
\providecommand \natexlab [1]{#1}%
\providecommand \enquote  [1]{``#1''}%
\providecommand \bibnamefont  [1]{#1}%
\providecommand \bibfnamefont [1]{#1}%
\providecommand \citenamefont [1]{#1}%
\providecommand \href@noop [0]{\@secondoftwo}%
\providecommand \href [0]{\begingroup \@sanitize@url \@href}%
\providecommand \@href[1]{\@@startlink{#1}\@@href}%
\providecommand \@@href[1]{\endgroup#1\@@endlink}%
\providecommand \@sanitize@url [0]{\catcode `\\12\catcode `\$12\catcode
  `\&12\catcode `\#12\catcode `\^12\catcode `\_12\catcode `\%12\relax}%
\providecommand \@@startlink[1]{}%
\providecommand \@@endlink[0]{}%
\providecommand \url  [0]{\begingroup\@sanitize@url \@url }%
\providecommand \@url [1]{\endgroup\@href {#1}{\urlprefix }}%
\providecommand \urlprefix  [0]{URL }%
\providecommand \Eprint [0]{\href }%
\providecommand \doibase [0]{http://dx.doi.org/}%
\providecommand \selectlanguage [0]{\@gobble}%
\providecommand \bibinfo  [0]{\@secondoftwo}%
\providecommand \bibfield  [0]{\@secondoftwo}%
\providecommand \translation [1]{[#1]}%
\providecommand \BibitemOpen [0]{}%
\providecommand \bibitemStop [0]{}%
\providecommand \bibitemNoStop [0]{.\EOS\space}%
\providecommand \EOS [0]{\spacefactor3000\relax}%
\providecommand \BibitemShut  [1]{\csname bibitem#1\endcsname}%
\let\auto@bib@innerbib\@empty
\bibitem [{\citenamefont {Custance}\ \emph {et~al.}(2009)\citenamefont
  {Custance}, \citenamefont {Perez},\ and\ \citenamefont
  {Morita}}]{Custance2009}%
  \BibitemOpen
  \bibfield  {author} {\bibinfo {author} {\bibfnamefont {Oscar}\ \bibnamefont
  {Custance}}, \bibinfo {author} {\bibfnamefont {Ruben}\ \bibnamefont {Perez}},
  \ and\ \bibinfo {author} {\bibfnamefont {Seizo}\ \bibnamefont {Morita}},\
  }\bibfield  {title} {\enquote {\bibinfo {title} {{Atomic force microscopy as
  a tool for atom manipulation}},}\ }\href {\doibase 10.1038/nnano.2009.347}
  {\bibfield  {journal} {\bibinfo  {journal} {Nat. Nanotechnol.}\ }\textbf
  {\bibinfo {volume} {4}},\ \bibinfo {pages} {803--810} (\bibinfo {year}
  {2009})}\BibitemShut {NoStop}%
\bibitem [{\citenamefont {Khajetoorians}\ \emph {et~al.}(2010)\citenamefont
  {Khajetoorians}, \citenamefont {Chilian}, \citenamefont {Wiebe},
  \citenamefont {Schuwalow}, \citenamefont {Lechermann},\ and\ \citenamefont
  {Wiesendanger}}]{Khajetoorians2010}%
  \BibitemOpen
  \bibfield  {author} {\bibinfo {author} {\bibfnamefont {Alexander~A.}\
  \bibnamefont {Khajetoorians}}, \bibinfo {author} {\bibfnamefont {Bruno}\
  \bibnamefont {Chilian}}, \bibinfo {author} {\bibfnamefont {Jens}\
  \bibnamefont {Wiebe}}, \bibinfo {author} {\bibfnamefont {Sergej}\
  \bibnamefont {Schuwalow}}, \bibinfo {author} {\bibfnamefont {Frank}\
  \bibnamefont {Lechermann}}, \ and\ \bibinfo {author} {\bibfnamefont {Roland}\
  \bibnamefont {Wiesendanger}},\ }\bibfield  {title} {\enquote {\bibinfo
  {title} {{Detecting excitation and magnetization of individual dopants in a
  semiconductor}},}\ }\href {\doibase 10.1038/nature09519} {\bibfield
  {journal} {\bibinfo  {journal} {Nature}\ }\textbf {\bibinfo {volume} {467}},\
  \bibinfo {pages} {1084--1087} (\bibinfo {year} {2010})}\BibitemShut {NoStop}%
\bibitem [{\citenamefont {Khajetoorians}\ \emph {et~al.}(2011)\citenamefont
  {Khajetoorians}, \citenamefont {Wiebe}, \citenamefont {Chilian},\ and\
  \citenamefont {Wiesendanger}}]{doi:10.1126/science.1201725}%
  \BibitemOpen
  \bibfield  {author} {\bibinfo {author} {\bibfnamefont {Alexander~Ako}\
  \bibnamefont {Khajetoorians}}, \bibinfo {author} {\bibfnamefont {Jens}\
  \bibnamefont {Wiebe}}, \bibinfo {author} {\bibfnamefont {Bruno}\ \bibnamefont
  {Chilian}}, \ and\ \bibinfo {author} {\bibfnamefont {Roland}\ \bibnamefont
  {Wiesendanger}},\ }\bibfield  {title} {\enquote {\bibinfo {title} {{Realizing
  All-Spin–Based Logic Operations Atom by Atom}},}\ }\href {\doibase
  10.1126/science.1201725} {\bibfield  {journal} {\bibinfo  {journal}
  {Science}\ }\textbf {\bibinfo {volume} {332}},\ \bibinfo {pages} {1062--1064}
  (\bibinfo {year} {2011})}\BibitemShut {NoStop}%
\bibitem [{\citenamefont {Khajetoorians}\ \emph {et~al.}(2012)\citenamefont
  {Khajetoorians}, \citenamefont {Wiebe}, \citenamefont {Chilian},
  \citenamefont {Lounis}, \citenamefont {Bl{\"u}gel},\ and\ \citenamefont
  {Wiesendanger}}]{Khajetoorians2012}%
  \BibitemOpen
  \bibfield  {author} {\bibinfo {author} {\bibfnamefont {Alexander~Ako}\
  \bibnamefont {Khajetoorians}}, \bibinfo {author} {\bibfnamefont {Jens}\
  \bibnamefont {Wiebe}}, \bibinfo {author} {\bibfnamefont {Bruno}\ \bibnamefont
  {Chilian}}, \bibinfo {author} {\bibfnamefont {Samir}\ \bibnamefont {Lounis}},
  \bibinfo {author} {\bibfnamefont {Stefan}\ \bibnamefont {Bl{\"u}gel}}, \ and\
  \bibinfo {author} {\bibfnamefont {Roland}\ \bibnamefont {Wiesendanger}},\
  }\bibfield  {title} {\enquote {\bibinfo {title} {{Atom-by-atom engineering
  and magnetometry of tailored nanomagnets}},}\ }\href {\doibase
  10.1038/nphys2299} {\bibfield  {journal} {\bibinfo  {journal} {Nat. Phys.}\
  }\textbf {\bibinfo {volume} {8}},\ \bibinfo {pages} {497--503} (\bibinfo
  {year} {2012})}\BibitemShut {NoStop}%
\bibitem [{\citenamefont {Schlenk}\ \emph {et~al.}(2013)\citenamefont
  {Schlenk}, \citenamefont {Bianchi}, \citenamefont {Koleini}, \citenamefont
  {Eich}, \citenamefont {Pietzsch}, \citenamefont {Wehling}, \citenamefont
  {Frauenheim}, \citenamefont {Balatsky}, \citenamefont {Mi}, \citenamefont
  {Iversen}, \citenamefont {Wiebe}, \citenamefont {Khajetoorians},
  \citenamefont {Hofmann},\ and\ \citenamefont
  {Wiesendanger}}]{PhysRevLett.110.126804}%
  \BibitemOpen
  \bibfield  {author} {\bibinfo {author} {\bibfnamefont {T.}~\bibnamefont
  {Schlenk}}, \bibinfo {author} {\bibfnamefont {M.}~\bibnamefont {Bianchi}},
  \bibinfo {author} {\bibfnamefont {M.}~\bibnamefont {Koleini}}, \bibinfo
  {author} {\bibfnamefont {A.}~\bibnamefont {Eich}}, \bibinfo {author}
  {\bibfnamefont {O.}~\bibnamefont {Pietzsch}}, \bibinfo {author}
  {\bibfnamefont {T.~O.}\ \bibnamefont {Wehling}}, \bibinfo {author}
  {\bibfnamefont {T.}~\bibnamefont {Frauenheim}}, \bibinfo {author}
  {\bibfnamefont {A.}~\bibnamefont {Balatsky}}, \bibinfo {author}
  {\bibfnamefont {J.-L.}\ \bibnamefont {Mi}}, \bibinfo {author} {\bibfnamefont
  {B.~B.}\ \bibnamefont {Iversen}}, \bibinfo {author} {\bibfnamefont
  {J.}~\bibnamefont {Wiebe}}, \bibinfo {author} {\bibfnamefont {A.~A.}\
  \bibnamefont {Khajetoorians}}, \bibinfo {author} {\bibfnamefont {Ph.}\
  \bibnamefont {Hofmann}}, \ and\ \bibinfo {author} {\bibfnamefont
  {R.}~\bibnamefont {Wiesendanger}},\ }\bibfield  {title} {\enquote {\bibinfo
  {title} {{Controllable Magnetic Doping of the Surface State of a Topological
  Insulator}},}\ }\href {\doibase 10.1103/PhysRevLett.110.126804} {\bibfield
  {journal} {\bibinfo  {journal} {Phys. Rev. Lett.}\ }\textbf {\bibinfo
  {volume} {110}},\ \bibinfo {pages} {126804} (\bibinfo {year}
  {2013})}\BibitemShut {NoStop}%
\bibitem [{\citenamefont {Khajetoorians}\ \emph {et~al.}(2013)\citenamefont
  {Khajetoorians}, \citenamefont {Schlenk}, \citenamefont {Schweflinghaus},
  \citenamefont {dos Santos~Dias}, \citenamefont {Steinbrecher}, \citenamefont
  {Bouhassoune}, \citenamefont {Lounis}, \citenamefont {Wiebe},\ and\
  \citenamefont {Wiesendanger}}]{PhysRevLett.111.157204}%
  \BibitemOpen
  \bibfield  {author} {\bibinfo {author} {\bibfnamefont {A.~A.}\ \bibnamefont
  {Khajetoorians}}, \bibinfo {author} {\bibfnamefont {T.}~\bibnamefont
  {Schlenk}}, \bibinfo {author} {\bibfnamefont {B.}~\bibnamefont
  {Schweflinghaus}}, \bibinfo {author} {\bibfnamefont {M.}~\bibnamefont {dos
  Santos~Dias}}, \bibinfo {author} {\bibfnamefont {M.}~\bibnamefont
  {Steinbrecher}}, \bibinfo {author} {\bibfnamefont {M.}~\bibnamefont
  {Bouhassoune}}, \bibinfo {author} {\bibfnamefont {S.}~\bibnamefont {Lounis}},
  \bibinfo {author} {\bibfnamefont {J.}~\bibnamefont {Wiebe}}, \ and\ \bibinfo
  {author} {\bibfnamefont {R.}~\bibnamefont {Wiesendanger}},\ }\bibfield
  {title} {\enquote {\bibinfo {title} {{Spin Excitations of Individual Fe Atoms
  on Pt(111): Impact of the Site-Dependent Giant Substrate Polarization}},}\
  }\href {\doibase 10.1103/PhysRevLett.111.157204} {\bibfield  {journal}
  {\bibinfo  {journal} {Phys. Rev. Lett.}\ }\textbf {\bibinfo {volume} {111}},\
  \bibinfo {pages} {157204} (\bibinfo {year} {2013})}\BibitemShut {NoStop}%
\bibitem [{\citenamefont {González-Herrero}\ \emph {et~al.}(2016)\citenamefont
  {González-Herrero}, \citenamefont {Gómez-Rodríguez}, \citenamefont
  {Mallet}, \citenamefont {Moaied}, \citenamefont {Palacios}, \citenamefont
  {Salgado}, \citenamefont {Ugeda}, \citenamefont {Veuillen}, \citenamefont
  {Yndurain},\ and\ \citenamefont {Brihuega}}]{doi:10.1126/science.aad8038}%
  \BibitemOpen
  \bibfield  {author} {\bibinfo {author} {\bibfnamefont {Héctor}\ \bibnamefont
  {González-Herrero}}, \bibinfo {author} {\bibfnamefont {José~M.}\
  \bibnamefont {Gómez-Rodríguez}}, \bibinfo {author} {\bibfnamefont {Pierre}\
  \bibnamefont {Mallet}}, \bibinfo {author} {\bibfnamefont {Mohamed}\
  \bibnamefont {Moaied}}, \bibinfo {author} {\bibfnamefont {Juan~José}\
  \bibnamefont {Palacios}}, \bibinfo {author} {\bibfnamefont {Carlos}\
  \bibnamefont {Salgado}}, \bibinfo {author} {\bibfnamefont {Miguel~M.}\
  \bibnamefont {Ugeda}}, \bibinfo {author} {\bibfnamefont {Jean-Yves}\
  \bibnamefont {Veuillen}}, \bibinfo {author} {\bibfnamefont {Félix}\
  \bibnamefont {Yndurain}}, \ and\ \bibinfo {author} {\bibfnamefont {Iván}\
  \bibnamefont {Brihuega}},\ }\bibfield  {title} {\enquote {\bibinfo {title}
  {{Atomic-scale control of graphene magnetism by using hydrogen atoms}},}\
  }\href {\doibase 10.1126/science.aad8038} {\bibfield  {journal} {\bibinfo
  {journal} {Science}\ }\textbf {\bibinfo {volume} {352}},\ \bibinfo {pages}
  {437--441} (\bibinfo {year} {2016})}\BibitemShut {NoStop}%
\bibitem [{\citenamefont {Khajetoorians}\ \emph {et~al.}(2016)\citenamefont
  {Khajetoorians}, \citenamefont {Steinbrecher}, \citenamefont {Ternes},
  \citenamefont {Bouhassoune}, \citenamefont {dos Santos~Dias}, \citenamefont
  {Lounis}, \citenamefont {Wiebe},\ and\ \citenamefont
  {Wiesendanger}}]{Khajetoorians2016}%
  \BibitemOpen
  \bibfield  {author} {\bibinfo {author} {\bibfnamefont {A.~A.}\ \bibnamefont
  {Khajetoorians}}, \bibinfo {author} {\bibfnamefont {M.}~\bibnamefont
  {Steinbrecher}}, \bibinfo {author} {\bibfnamefont {M.}~\bibnamefont
  {Ternes}}, \bibinfo {author} {\bibfnamefont {M.}~\bibnamefont {Bouhassoune}},
  \bibinfo {author} {\bibfnamefont {M.}~\bibnamefont {dos Santos~Dias}},
  \bibinfo {author} {\bibfnamefont {S.}~\bibnamefont {Lounis}}, \bibinfo
  {author} {\bibfnamefont {J.}~\bibnamefont {Wiebe}}, \ and\ \bibinfo {author}
  {\bibfnamefont {R.}~\bibnamefont {Wiesendanger}},\ }\bibfield  {title}
  {\enquote {\bibinfo {title} {{Tailoring the chiral magnetic interaction
  between two individual atoms}},}\ }\href {\doibase 10.1038/ncomms10620}
  {\bibfield  {journal} {\bibinfo  {journal} {Nat. Commun.}\ }\textbf {\bibinfo
  {volume} {7}},\ \bibinfo {pages} {10620} (\bibinfo {year}
  {2016})}\BibitemShut {NoStop}%
\bibitem [{\citenamefont {Slez\'ak}\ \emph {et~al.}(1999)\citenamefont
  {Slez\'ak}, \citenamefont {Mutombo},\ and\ \citenamefont
  {Ch\'ab}}]{PhysRevB.60.13328}%
  \BibitemOpen
  \bibfield  {author} {\bibinfo {author} {\bibfnamefont {J.}~\bibnamefont
  {Slez\'ak}}, \bibinfo {author} {\bibfnamefont {P.}~\bibnamefont {Mutombo}}, \
  and\ \bibinfo {author} {\bibfnamefont {V.}~\bibnamefont {Ch\'ab}},\
  }\bibfield  {title} {\enquote {\bibinfo {title} {{STM study of a Pb/Si(111)
  interface at room and low temperatures}},}\ }\href {\doibase
  10.1103/PhysRevB.60.13328} {\bibfield  {journal} {\bibinfo  {journal} {Phys.
  Rev. B}\ }\textbf {\bibinfo {volume} {60}},\ \bibinfo {pages} {13328--13330}
  (\bibinfo {year} {1999})}\BibitemShut {NoStop}%
\bibitem [{\citenamefont {Lobo}\ \emph {et~al.}(2003)\citenamefont {Lobo},
  \citenamefont {Tejeda}, \citenamefont {Mugarza},\ and\ \citenamefont
  {Michel}}]{PhysRevB.68.235332}%
  \BibitemOpen
  \bibfield  {author} {\bibinfo {author} {\bibfnamefont {J.}~\bibnamefont
  {Lobo}}, \bibinfo {author} {\bibfnamefont {A.}~\bibnamefont {Tejeda}},
  \bibinfo {author} {\bibfnamefont {A.}~\bibnamefont {Mugarza}}, \ and\
  \bibinfo {author} {\bibfnamefont {E.~G.}\ \bibnamefont {Michel}},\ }\bibfield
   {title} {\enquote {\bibinfo {title} {{Electronic structure of
  Sn/Si(111)-$(\sqrt{3}\ifmmode\times\else\texttimes\fi{}\sqrt{3})R30\ifmmode^\circ\else\textdegree\fi{}$
  as a function of Sn coverage}},}\ }\href {\doibase
  10.1103/PhysRevB.68.235332} {\bibfield  {journal} {\bibinfo  {journal} {Phys.
  Rev. B}\ }\textbf {\bibinfo {volume} {68}},\ \bibinfo {pages} {235332}
  (\bibinfo {year} {2003})}\BibitemShut {NoStop}%
\bibitem [{\citenamefont {Upton}\ \emph {et~al.}(2005)\citenamefont {Upton},
  \citenamefont {Miller},\ and\ \citenamefont {Chiang}}]{PhysRevB.71.033403}%
  \BibitemOpen
  \bibfield  {author} {\bibinfo {author} {\bibfnamefont {M.~H.}\ \bibnamefont
  {Upton}}, \bibinfo {author} {\bibfnamefont {T.}~\bibnamefont {Miller}}, \
  and\ \bibinfo {author} {\bibfnamefont {T.-C.}\ \bibnamefont {Chiang}},\
  }\bibfield  {title} {\enquote {\bibinfo {title} {{Unusual band dispersion in
  Pb films on Si(111)}},}\ }\href {\doibase 10.1103/PhysRevB.71.033403}
  {\bibfield  {journal} {\bibinfo  {journal} {Phys. Rev. B}\ }\textbf {\bibinfo
  {volume} {71}},\ \bibinfo {pages} {033403} (\bibinfo {year}
  {2005})}\BibitemShut {NoStop}%
\bibitem [{\citenamefont {Modesti}\ \emph {et~al.}(2007)\citenamefont
  {Modesti}, \citenamefont {Petaccia}, \citenamefont {Ceballos}, \citenamefont
  {Vobornik}, \citenamefont {Panaccione}, \citenamefont {Rossi}, \citenamefont
  {Ottaviano}, \citenamefont {Larciprete}, \citenamefont {Lizzit},\ and\
  \citenamefont {Goldoni}}]{PhysRevLett.98.126401}%
  \BibitemOpen
  \bibfield  {author} {\bibinfo {author} {\bibfnamefont {S.}~\bibnamefont
  {Modesti}}, \bibinfo {author} {\bibfnamefont {L.}~\bibnamefont {Petaccia}},
  \bibinfo {author} {\bibfnamefont {G.}~\bibnamefont {Ceballos}}, \bibinfo
  {author} {\bibfnamefont {I.}~\bibnamefont {Vobornik}}, \bibinfo {author}
  {\bibfnamefont {G.}~\bibnamefont {Panaccione}}, \bibinfo {author}
  {\bibfnamefont {G.}~\bibnamefont {Rossi}}, \bibinfo {author} {\bibfnamefont
  {L.}~\bibnamefont {Ottaviano}}, \bibinfo {author} {\bibfnamefont
  {R.}~\bibnamefont {Larciprete}}, \bibinfo {author} {\bibfnamefont
  {S.}~\bibnamefont {Lizzit}}, \ and\ \bibinfo {author} {\bibfnamefont
  {A.}~\bibnamefont {Goldoni}},\ }\bibfield  {title} {\enquote {\bibinfo
  {title} {{Insulating Ground State of
  $\mathrm{Sn}/\mathrm{Si}(111)\mathrm{\text{\ensuremath{-}}}(\sqrt{3}\ifmmode\times\else\texttimes\fi{}\sqrt{3})R30\ifmmode^\circ\else\textdegree\fi{}$}},}\
  }\href {\doibase 10.1103/PhysRevLett.98.126401} {\bibfield  {journal}
  {\bibinfo  {journal} {Phys. Rev. Lett.}\ }\textbf {\bibinfo {volume} {98}},\
  \bibinfo {pages} {126401} (\bibinfo {year} {2007})}\BibitemShut {NoStop}%
\bibitem [{\citenamefont {Zhang}\ \emph {et~al.}(2010)\citenamefont {Zhang},
  \citenamefont {Cheng}, \citenamefont {Li}, \citenamefont {Sun}, \citenamefont
  {Wang}, \citenamefont {Zhu}, \citenamefont {He}, \citenamefont {Wang},
  \citenamefont {Ma}, \citenamefont {Chen}, \citenamefont {Wang}, \citenamefont
  {Liu}, \citenamefont {Lin}, \citenamefont {Jia},\ and\ \citenamefont
  {Xue}}]{Zhang2010}%
  \BibitemOpen
  \bibfield  {author} {\bibinfo {author} {\bibfnamefont {Tong}\ \bibnamefont
  {Zhang}}, \bibinfo {author} {\bibfnamefont {Peng}\ \bibnamefont {Cheng}},
  \bibinfo {author} {\bibfnamefont {Wen-Juan}\ \bibnamefont {Li}}, \bibinfo
  {author} {\bibfnamefont {Yu-Jie}\ \bibnamefont {Sun}}, \bibinfo {author}
  {\bibfnamefont {Guang}\ \bibnamefont {Wang}}, \bibinfo {author}
  {\bibfnamefont {Xie-Gang}\ \bibnamefont {Zhu}}, \bibinfo {author}
  {\bibfnamefont {Ke}~\bibnamefont {He}}, \bibinfo {author} {\bibfnamefont
  {Lili}\ \bibnamefont {Wang}}, \bibinfo {author} {\bibfnamefont {Xucun}\
  \bibnamefont {Ma}}, \bibinfo {author} {\bibfnamefont {Xi}~\bibnamefont
  {Chen}}, \bibinfo {author} {\bibfnamefont {Yayu}\ \bibnamefont {Wang}},
  \bibinfo {author} {\bibfnamefont {Ying}\ \bibnamefont {Liu}}, \bibinfo
  {author} {\bibfnamefont {Hai-Qing}\ \bibnamefont {Lin}}, \bibinfo {author}
  {\bibfnamefont {Jin-Feng}\ \bibnamefont {Jia}}, \ and\ \bibinfo {author}
  {\bibfnamefont {Qi-Kun}\ \bibnamefont {Xue}},\ }\bibfield  {title} {\enquote
  {\bibinfo {title} {{Superconductivity in one-atomic-layer metal films grown
  on Si(111)}},}\ }\href {\doibase 10.1038/nphys1499} {\bibfield  {journal}
  {\bibinfo  {journal} {Nat. Phys.}\ }\textbf {\bibinfo {volume} {6}},\
  \bibinfo {pages} {104--108} (\bibinfo {year} {2010})}\BibitemShut {NoStop}%
\bibitem [{\citenamefont {Li}\ \emph {et~al.}(2013)\citenamefont {Li},
  \citenamefont {H{\"o}pfner}, \citenamefont {Sch{\"a}fer}, \citenamefont
  {Blumenstein}, \citenamefont {Meyer}, \citenamefont {Bostwick}, \citenamefont
  {Rotenberg}, \citenamefont {Claessen},\ and\ \citenamefont {Hanke}}]{Li2013}%
  \BibitemOpen
  \bibfield  {author} {\bibinfo {author} {\bibfnamefont {Gang}\ \bibnamefont
  {Li}}, \bibinfo {author} {\bibfnamefont {Philipp}\ \bibnamefont
  {H{\"o}pfner}}, \bibinfo {author} {\bibfnamefont {J{\"o}rg}\ \bibnamefont
  {Sch{\"a}fer}}, \bibinfo {author} {\bibfnamefont {Christian}\ \bibnamefont
  {Blumenstein}}, \bibinfo {author} {\bibfnamefont {Sebastian}\ \bibnamefont
  {Meyer}}, \bibinfo {author} {\bibfnamefont {Aaron}\ \bibnamefont {Bostwick}},
  \bibinfo {author} {\bibfnamefont {Eli}\ \bibnamefont {Rotenberg}}, \bibinfo
  {author} {\bibfnamefont {Ralph}\ \bibnamefont {Claessen}}, \ and\ \bibinfo
  {author} {\bibfnamefont {Werner}\ \bibnamefont {Hanke}},\ }\bibfield  {title}
  {\enquote {\bibinfo {title} {{Magnetic order in a frustrated two-dimensional
  atom lattice at a semiconductor surface}},}\ }\href {\doibase
  10.1038/ncomms2617} {\bibfield  {journal} {\bibinfo  {journal} {Nat.
  Commun.}\ }\textbf {\bibinfo {volume} {4}},\ \bibinfo {pages} {1620}
  (\bibinfo {year} {2013})}\BibitemShut {NoStop}%
\bibitem [{\citenamefont {Tresca}\ \emph {et~al.}(2018)\citenamefont {Tresca},
  \citenamefont {Brun}, \citenamefont {Bilgeri}, \citenamefont {Menard},
  \citenamefont {Cherkez}, \citenamefont {Federicci}, \citenamefont {Longo},
  \citenamefont {Debontridder}, \citenamefont {D'angelo}, \citenamefont
  {Roditchev}, \citenamefont {Profeta}, \citenamefont {Calandra},\ and\
  \citenamefont {Cren}}]{PhysRevLett.120.196402}%
  \BibitemOpen
  \bibfield  {author} {\bibinfo {author} {\bibfnamefont {C.}~\bibnamefont
  {Tresca}}, \bibinfo {author} {\bibfnamefont {C.}~\bibnamefont {Brun}},
  \bibinfo {author} {\bibfnamefont {T.}~\bibnamefont {Bilgeri}}, \bibinfo
  {author} {\bibfnamefont {G.}~\bibnamefont {Menard}}, \bibinfo {author}
  {\bibfnamefont {V.}~\bibnamefont {Cherkez}}, \bibinfo {author} {\bibfnamefont
  {R.}~\bibnamefont {Federicci}}, \bibinfo {author} {\bibfnamefont
  {D.}~\bibnamefont {Longo}}, \bibinfo {author} {\bibfnamefont
  {F.}~\bibnamefont {Debontridder}}, \bibinfo {author} {\bibfnamefont
  {M.}~\bibnamefont {D'angelo}}, \bibinfo {author} {\bibfnamefont
  {D.}~\bibnamefont {Roditchev}}, \bibinfo {author} {\bibfnamefont
  {G.}~\bibnamefont {Profeta}}, \bibinfo {author} {\bibfnamefont
  {M.}~\bibnamefont {Calandra}}, \ and\ \bibinfo {author} {\bibfnamefont
  {T.}~\bibnamefont {Cren}},\ }\bibfield  {title} {\enquote {\bibinfo {title}
  {{Chiral Spin Texture in the Charge-Density-Wave Phase of the Correlated
  Metallic $\mathrm{Pb}/\mathrm{Si}(111)$ Monolayer}},}\ }\href {\doibase
  10.1103/PhysRevLett.120.196402} {\bibfield  {journal} {\bibinfo  {journal}
  {Phys. Rev. Lett.}\ }\textbf {\bibinfo {volume} {120}},\ \bibinfo {pages}
  {196402} (\bibinfo {year} {2018})}\BibitemShut {NoStop}%
\bibitem [{\citenamefont {Carpinelli}\ \emph {et~al.}(1997)\citenamefont
  {Carpinelli}, \citenamefont {Weitering}, \citenamefont {Bartkowiak},
  \citenamefont {Stumpf},\ and\ \citenamefont {Plummer}}]{PhysRevLett.79.2859}%
  \BibitemOpen
  \bibfield  {author} {\bibinfo {author} {\bibfnamefont {J.~M.}\ \bibnamefont
  {Carpinelli}}, \bibinfo {author} {\bibfnamefont {H.~H.}\ \bibnamefont
  {Weitering}}, \bibinfo {author} {\bibfnamefont {M.}~\bibnamefont
  {Bartkowiak}}, \bibinfo {author} {\bibfnamefont {R.}~\bibnamefont {Stumpf}},
  \ and\ \bibinfo {author} {\bibfnamefont {E.~W.}\ \bibnamefont {Plummer}},\
  }\bibfield  {title} {\enquote {\bibinfo {title} {Surface charge ordering
  transition: $\mathit{\ensuremath{\alpha}}$ phase of sn/ge(111)},}\ }\href
  {\doibase 10.1103/PhysRevLett.79.2859} {\bibfield  {journal} {\bibinfo
  {journal} {Phys. Rev. Lett.}\ }\textbf {\bibinfo {volume} {79}},\ \bibinfo
  {pages} {2859--2862} (\bibinfo {year} {1997})}\BibitemShut {NoStop}%
\bibitem [{\citenamefont {Floreano}\ \emph {et~al.}(2001)\citenamefont
  {Floreano}, \citenamefont {Cvetko}, \citenamefont {Bavdek}, \citenamefont
  {Benes},\ and\ \citenamefont {Morgante}}]{PhysRevB.64.075405}%
  \BibitemOpen
  \bibfield  {author} {\bibinfo {author} {\bibfnamefont {L.}~\bibnamefont
  {Floreano}}, \bibinfo {author} {\bibfnamefont {D.}~\bibnamefont {Cvetko}},
  \bibinfo {author} {\bibfnamefont {G.}~\bibnamefont {Bavdek}}, \bibinfo
  {author} {\bibfnamefont {M.}~\bibnamefont {Benes}}, \ and\ \bibinfo {author}
  {\bibfnamefont {A.}~\bibnamefont {Morgante}},\ }\bibfield  {title} {\enquote
  {\bibinfo {title} {{Order-disorder transition of the
  $(3\ifmmode\times\else\texttimes\fi{}3)$ Sn/Ge(111) phase}},}\ }\href
  {\doibase 10.1103/PhysRevB.64.075405} {\bibfield  {journal} {\bibinfo
  {journal} {Phys. Rev. B}\ }\textbf {\bibinfo {volume} {64}},\ \bibinfo
  {pages} {075405} (\bibinfo {year} {2001})}\BibitemShut {NoStop}%
\bibitem [{\citenamefont {Tresca}\ and\ \citenamefont
  {Calandra}(2021)}]{PhysRevB.104.045126}%
  \BibitemOpen
  \bibfield  {author} {\bibinfo {author} {\bibfnamefont {Cesare}\ \bibnamefont
  {Tresca}}\ and\ \bibinfo {author} {\bibfnamefont {Matteo}\ \bibnamefont
  {Calandra}},\ }\bibfield  {title} {\enquote {\bibinfo {title} {{Charge
  density wave in single-layer Pb/Ge(111) driven by Pb-substrate exchange
  interaction}},}\ }\href {\doibase 10.1103/PhysRevB.104.045126} {\bibfield
  {journal} {\bibinfo  {journal} {Phys. Rev. B}\ }\textbf {\bibinfo {volume}
  {104}},\ \bibinfo {pages} {045126} (\bibinfo {year} {2021})}\BibitemShut
  {NoStop}%
\bibitem [{\citenamefont {Glass}\ \emph {et~al.}(2015)\citenamefont {Glass},
  \citenamefont {Li}, \citenamefont {Adler}, \citenamefont {Aulbach},
  \citenamefont {Fleszar}, \citenamefont {Thomale}, \citenamefont {Hanke},
  \citenamefont {Claessen},\ and\ \citenamefont
  {Sch\"afer}}]{PhysRevLett.114.247602}%
  \BibitemOpen
  \bibfield  {author} {\bibinfo {author} {\bibfnamefont {S.}~\bibnamefont
  {Glass}}, \bibinfo {author} {\bibfnamefont {G.}~\bibnamefont {Li}}, \bibinfo
  {author} {\bibfnamefont {F.}~\bibnamefont {Adler}}, \bibinfo {author}
  {\bibfnamefont {J.}~\bibnamefont {Aulbach}}, \bibinfo {author} {\bibfnamefont
  {A.}~\bibnamefont {Fleszar}}, \bibinfo {author} {\bibfnamefont
  {R.}~\bibnamefont {Thomale}}, \bibinfo {author} {\bibfnamefont
  {W.}~\bibnamefont {Hanke}}, \bibinfo {author} {\bibfnamefont
  {R.}~\bibnamefont {Claessen}}, \ and\ \bibinfo {author} {\bibfnamefont
  {J.}~\bibnamefont {Sch\"afer}},\ }\bibfield  {title} {\enquote {\bibinfo
  {title} {{Triangular Spin-Orbit-Coupled Lattice with Strong Coulomb
  Correlations: Sn Atoms on a SiC(0001) Substrate}},}\ }\href {\doibase
  10.1103/PhysRevLett.114.247602} {\bibfield  {journal} {\bibinfo  {journal}
  {Phys. Rev. Lett.}\ }\textbf {\bibinfo {volume} {114}},\ \bibinfo {pages}
  {247602} (\bibinfo {year} {2015})}\BibitemShut {NoStop}%
\bibitem [{\citenamefont {Kennes}\ \emph {et~al.}(2021)\citenamefont {Kennes},
  \citenamefont {Claassen}, \citenamefont {Xian}, \citenamefont {Georges},
  \citenamefont {Millis}, \citenamefont {Hone}, \citenamefont {Dean},
  \citenamefont {Basov}, \citenamefont {Pasupathy},\ and\ \citenamefont
  {Rubio}}]{Kennes2021}%
  \BibitemOpen
  \bibfield  {author} {\bibinfo {author} {\bibfnamefont {Dante~M.}\
  \bibnamefont {Kennes}}, \bibinfo {author} {\bibfnamefont {Martin}\
  \bibnamefont {Claassen}}, \bibinfo {author} {\bibfnamefont {Lede}\
  \bibnamefont {Xian}}, \bibinfo {author} {\bibfnamefont {Antoine}\
  \bibnamefont {Georges}}, \bibinfo {author} {\bibfnamefont {Andrew~J.}\
  \bibnamefont {Millis}}, \bibinfo {author} {\bibfnamefont {James}\
  \bibnamefont {Hone}}, \bibinfo {author} {\bibfnamefont {Cory~R.}\
  \bibnamefont {Dean}}, \bibinfo {author} {\bibfnamefont {D.~N.}\ \bibnamefont
  {Basov}}, \bibinfo {author} {\bibfnamefont {Abhay~N.}\ \bibnamefont
  {Pasupathy}}, \ and\ \bibinfo {author} {\bibfnamefont {Angel}\ \bibnamefont
  {Rubio}},\ }\bibfield  {title} {\enquote {\bibinfo {title} {Moir{\'e}
  heterostructures as a condensed-matter quantum simulator},}\ }\href {\doibase
  10.1038/s41567-020-01154-3} {\bibfield  {journal} {\bibinfo  {journal} {Nat.
  Phys.}\ }\textbf {\bibinfo {volume} {17}},\ \bibinfo {pages} {155--163}
  (\bibinfo {year} {2021})}\BibitemShut {NoStop}%
\bibitem [{\citenamefont {Polini}\ \emph {et~al.}(2013)\citenamefont {Polini},
  \citenamefont {Guinea}, \citenamefont {Lewenstein}, \citenamefont
  {Manoharan},\ and\ \citenamefont {Pellegrini}}]{Polini2013}%
  \BibitemOpen
  \bibfield  {author} {\bibinfo {author} {\bibfnamefont {Marco}\ \bibnamefont
  {Polini}}, \bibinfo {author} {\bibfnamefont {Francisco}\ \bibnamefont
  {Guinea}}, \bibinfo {author} {\bibfnamefont {Maciej}\ \bibnamefont
  {Lewenstein}}, \bibinfo {author} {\bibfnamefont {Hari~C.}\ \bibnamefont
  {Manoharan}}, \ and\ \bibinfo {author} {\bibfnamefont {Vittorio}\
  \bibnamefont {Pellegrini}},\ }\bibfield  {title} {\enquote {\bibinfo {title}
  {{Artificial honeycomb lattices for electrons, atoms and photons}},}\ }\href
  {\doibase 10.1038/nnano.2013.161} {\bibfield  {journal} {\bibinfo  {journal}
  {Nat. Nanotechnol.}\ }\textbf {\bibinfo {volume} {8}},\ \bibinfo {pages}
  {625--633} (\bibinfo {year} {2013})}\BibitemShut {NoStop}%
\bibitem [{\citenamefont {Khajetoorians}\ \emph {et~al.}(2019)\citenamefont
  {Khajetoorians}, \citenamefont {Wegner}, \citenamefont {Otte},\ and\
  \citenamefont {Swart}}]{Khajetoorians2019}%
  \BibitemOpen
  \bibfield  {author} {\bibinfo {author} {\bibfnamefont {Alexander~A.}\
  \bibnamefont {Khajetoorians}}, \bibinfo {author} {\bibfnamefont {Daniel}\
  \bibnamefont {Wegner}}, \bibinfo {author} {\bibfnamefont {Alexander~F.}\
  \bibnamefont {Otte}}, \ and\ \bibinfo {author} {\bibfnamefont {Ingmar}\
  \bibnamefont {Swart}},\ }\bibfield  {title} {\enquote {\bibinfo {title}
  {{Creating designer quantum states of matter atom-by-atom}},}\ }\href
  {\doibase 10.1038/s42254-019-0108-5} {\bibfield  {journal} {\bibinfo
  {journal} {Nat. Rev. Phys.}\ }\textbf {\bibinfo {volume} {1}},\ \bibinfo
  {pages} {703--715} (\bibinfo {year} {2019})}\BibitemShut {NoStop}%
\bibitem [{\citenamefont {Wolf}\ \emph {et~al.}(2022)\citenamefont {Wolf},
  \citenamefont {Di~Sante}, \citenamefont {Schwemmer}, \citenamefont
  {Thomale},\ and\ \citenamefont {Rachel}}]{PhysRevLett.128.167002}%
  \BibitemOpen
  \bibfield  {author} {\bibinfo {author} {\bibfnamefont {Sebastian}\
  \bibnamefont {Wolf}}, \bibinfo {author} {\bibfnamefont {Domenico}\
  \bibnamefont {Di~Sante}}, \bibinfo {author} {\bibfnamefont {Tilman}\
  \bibnamefont {Schwemmer}}, \bibinfo {author} {\bibfnamefont {Ronny}\
  \bibnamefont {Thomale}}, \ and\ \bibinfo {author} {\bibfnamefont {Stephan}\
  \bibnamefont {Rachel}},\ }\bibfield  {title} {\enquote {\bibinfo {title}
  {{Triplet Superconductivity from Nonlocal Coulomb Repulsion in an Atomic Sn
  Layer Deposited onto a Si(111) Substrate}},}\ }\href {\doibase
  10.1103/PhysRevLett.128.167002} {\bibfield  {journal} {\bibinfo  {journal}
  {Phys. Rev. Lett.}\ }\textbf {\bibinfo {volume} {128}},\ \bibinfo {pages}
  {167002} (\bibinfo {year} {2022})}\BibitemShut {NoStop}%
\bibitem [{\citenamefont {Rohringer}\ \emph {et~al.}(2018)\citenamefont
  {Rohringer}, \citenamefont {Hafermann}, \citenamefont {Toschi}, \citenamefont
  {Katanin}, \citenamefont {Antipov}, \citenamefont {Katsnelson}, \citenamefont
  {Lichtenstein}, \citenamefont {Rubtsov},\ and\ \citenamefont
  {Held}}]{RevModPhys.90.025003}%
  \BibitemOpen
  \bibfield  {author} {\bibinfo {author} {\bibfnamefont {G.}~\bibnamefont
  {Rohringer}}, \bibinfo {author} {\bibfnamefont {H.}~\bibnamefont
  {Hafermann}}, \bibinfo {author} {\bibfnamefont {A.}~\bibnamefont {Toschi}},
  \bibinfo {author} {\bibfnamefont {A.~A.}\ \bibnamefont {Katanin}}, \bibinfo
  {author} {\bibfnamefont {A.~E.}\ \bibnamefont {Antipov}}, \bibinfo {author}
  {\bibfnamefont {M.~I.}\ \bibnamefont {Katsnelson}}, \bibinfo {author}
  {\bibfnamefont {A.~I.}\ \bibnamefont {Lichtenstein}}, \bibinfo {author}
  {\bibfnamefont {A.~N.}\ \bibnamefont {Rubtsov}}, \ and\ \bibinfo {author}
  {\bibfnamefont {K.}~\bibnamefont {Held}},\ }\bibfield  {title} {\enquote
  {\bibinfo {title} {Diagrammatic routes to nonlocal correlations beyond
  dynamical mean field theory},}\ }\href {\doibase
  10.1103/RevModPhys.90.025003} {\bibfield  {journal} {\bibinfo  {journal}
  {Rev. Mod. Phys.}\ }\textbf {\bibinfo {volume} {90}},\ \bibinfo {pages}
  {025003} (\bibinfo {year} {2018})}\BibitemShut {NoStop}%
\bibitem [{\citenamefont {Sch\"afer}\ \emph {et~al.}(2021)\citenamefont
  {Sch\"afer}, \citenamefont {Wentzell}, \citenamefont {\ifmmode~\check{S}\else
  \v{S}\fi{}imkovic}, \citenamefont {He}, \citenamefont {Hille}, \citenamefont
  {Klett}, \citenamefont {Eckhardt}, \citenamefont {Arzhang}, \citenamefont
  {Harkov}, \citenamefont {Le~R\'egent}, \citenamefont {Kirsch}, \citenamefont
  {Wang}, \citenamefont {Kim}, \citenamefont {Kozik}, \citenamefont {Stepanov},
  \citenamefont {Kauch}, \citenamefont {Andergassen}, \citenamefont {Hansmann},
  \citenamefont {Rohe}, \citenamefont {Vilk}, \citenamefont {LeBlanc},
  \citenamefont {Zhang}, \citenamefont {Tremblay}, \citenamefont {Ferrero},
  \citenamefont {Parcollet},\ and\ \citenamefont
  {Georges}}]{PhysRevX.11.011058}%
  \BibitemOpen
  \bibfield  {author} {\bibinfo {author} {\bibfnamefont {Thomas}\ \bibnamefont
  {Sch\"afer}}, \bibinfo {author} {\bibfnamefont {Nils}\ \bibnamefont
  {Wentzell}}, \bibinfo {author} {\bibfnamefont {Fedor}\ \bibnamefont
  {\ifmmode~\check{S}\else \v{S}\fi{}imkovic}}, \bibinfo {author}
  {\bibfnamefont {Yuan-Yao}\ \bibnamefont {He}}, \bibinfo {author}
  {\bibfnamefont {Cornelia}\ \bibnamefont {Hille}}, \bibinfo {author}
  {\bibfnamefont {Marcel}\ \bibnamefont {Klett}}, \bibinfo {author}
  {\bibfnamefont {Christian~J.}\ \bibnamefont {Eckhardt}}, \bibinfo {author}
  {\bibfnamefont {Behnam}\ \bibnamefont {Arzhang}}, \bibinfo {author}
  {\bibfnamefont {Viktor}\ \bibnamefont {Harkov}}, \bibinfo {author}
  {\bibfnamefont {Fran\ifmmode \mbox{\c{c}}\else \c{c}\fi{}ois-Marie}\
  \bibnamefont {Le~R\'egent}}, \bibinfo {author} {\bibfnamefont {Alfred}\
  \bibnamefont {Kirsch}}, \bibinfo {author} {\bibfnamefont {Yan}\ \bibnamefont
  {Wang}}, \bibinfo {author} {\bibfnamefont {Aaram~J.}\ \bibnamefont {Kim}},
  \bibinfo {author} {\bibfnamefont {Evgeny}\ \bibnamefont {Kozik}}, \bibinfo
  {author} {\bibfnamefont {Evgeny~A.}\ \bibnamefont {Stepanov}}, \bibinfo
  {author} {\bibfnamefont {Anna}\ \bibnamefont {Kauch}}, \bibinfo {author}
  {\bibfnamefont {Sabine}\ \bibnamefont {Andergassen}}, \bibinfo {author}
  {\bibfnamefont {Philipp}\ \bibnamefont {Hansmann}}, \bibinfo {author}
  {\bibfnamefont {Daniel}\ \bibnamefont {Rohe}}, \bibinfo {author}
  {\bibfnamefont {Yuri~M.}\ \bibnamefont {Vilk}}, \bibinfo {author}
  {\bibfnamefont {James P.~F.}\ \bibnamefont {LeBlanc}}, \bibinfo {author}
  {\bibfnamefont {Shiwei}\ \bibnamefont {Zhang}}, \bibinfo {author}
  {\bibfnamefont {A.-M.~S.}\ \bibnamefont {Tremblay}}, \bibinfo {author}
  {\bibfnamefont {Michel}\ \bibnamefont {Ferrero}}, \bibinfo {author}
  {\bibfnamefont {Olivier}\ \bibnamefont {Parcollet}}, \ and\ \bibinfo {author}
  {\bibfnamefont {Antoine}\ \bibnamefont {Georges}},\ }\bibfield  {title}
  {\enquote {\bibinfo {title} {{Tracking the Footprints of Spin Fluctuations: A
  MultiMethod, MultiMessenger Study of the Two-Dimensional Hubbard Model}},}\
  }\href {\doibase 10.1103/PhysRevX.11.011058} {\bibfield  {journal} {\bibinfo
  {journal} {Phys. Rev. X}\ }\textbf {\bibinfo {volume} {11}},\ \bibinfo
  {pages} {011058} (\bibinfo {year} {2021})}\BibitemShut {NoStop}%
\bibitem [{\citenamefont {Hansmann}\ \emph
  {et~al.}(2013{\natexlab{a}})\citenamefont {Hansmann}, \citenamefont
  {Vaugier}, \citenamefont {Jiang},\ and\ \citenamefont
  {Biermann}}]{Hansmann_2013}%
  \BibitemOpen
  \bibfield  {author} {\bibinfo {author} {\bibfnamefont {Philipp}\ \bibnamefont
  {Hansmann}}, \bibinfo {author} {\bibfnamefont {Lo{\"i}g}\ \bibnamefont
  {Vaugier}}, \bibinfo {author} {\bibfnamefont {Hong}\ \bibnamefont {Jiang}}, \
  and\ \bibinfo {author} {\bibfnamefont {Silke}\ \bibnamefont {Biermann}},\
  }\bibfield  {title} {\enquote {\bibinfo {title} {{What about $U$ on surfaces?
  Extended Hubbard models for adatom systems from first principles}},}\ }\href
  {\doibase 10.1088/0953-8984/25/9/094005} {\bibfield  {journal} {\bibinfo
  {journal} {J. Phys. Condens. Matter}\ }\textbf {\bibinfo {volume} {25}},\
  \bibinfo {pages} {094005} (\bibinfo {year} {2013}{\natexlab{a}})}\BibitemShut
  {NoStop}%
\bibitem [{\citenamefont {Hansmann}\ \emph
  {et~al.}(2013{\natexlab{b}})\citenamefont {Hansmann}, \citenamefont {Ayral},
  \citenamefont {Vaugier}, \citenamefont {Werner},\ and\ \citenamefont
  {Biermann}}]{PhysRevLett.110.166401}%
  \BibitemOpen
  \bibfield  {author} {\bibinfo {author} {\bibfnamefont {P.}~\bibnamefont
  {Hansmann}}, \bibinfo {author} {\bibfnamefont {T.}~\bibnamefont {Ayral}},
  \bibinfo {author} {\bibfnamefont {L.}~\bibnamefont {Vaugier}}, \bibinfo
  {author} {\bibfnamefont {P.}~\bibnamefont {Werner}}, \ and\ \bibinfo {author}
  {\bibfnamefont {S.}~\bibnamefont {Biermann}},\ }\bibfield  {title} {\enquote
  {\bibinfo {title} {{Long-Range Coulomb Interactions in Surface Systems: A
  First-Principles Description within Self-Consistently Combined $GW$ and
  Dynamical Mean-Field Theory}},}\ }\href {\doibase
  10.1103/PhysRevLett.110.166401} {\bibfield  {journal} {\bibinfo  {journal}
  {Phys. Rev. Lett.}\ }\textbf {\bibinfo {volume} {110}},\ \bibinfo {pages}
  {166401} (\bibinfo {year} {2013}{\natexlab{b}})}\BibitemShut {NoStop}%
\bibitem [{\citenamefont {Badrtdinov}\ \emph {et~al.}(2016)\citenamefont
  {Badrtdinov}, \citenamefont {Nikolaev}, \citenamefont {Katsnelson},\ and\
  \citenamefont {Mazurenko}}]{PhysRevB.94.224418}%
  \BibitemOpen
  \bibfield  {author} {\bibinfo {author} {\bibfnamefont {D.~I.}\ \bibnamefont
  {Badrtdinov}}, \bibinfo {author} {\bibfnamefont {S.~A.}\ \bibnamefont
  {Nikolaev}}, \bibinfo {author} {\bibfnamefont {M.~I.}\ \bibnamefont
  {Katsnelson}}, \ and\ \bibinfo {author} {\bibfnamefont {V.~V.}\ \bibnamefont
  {Mazurenko}},\ }\bibfield  {title} {\enquote {\bibinfo {title} {{Spin-orbit
  coupling and magnetic interactions in Si(111):\{C,Si,Sn,Pb\}}},}\ }\href
  {\doibase 10.1103/PhysRevB.94.224418} {\bibfield  {journal} {\bibinfo
  {journal} {Phys. Rev. B}\ }\textbf {\bibinfo {volume} {94}},\ \bibinfo
  {pages} {224418} (\bibinfo {year} {2016})}\BibitemShut {NoStop}%
\bibitem [{\citenamefont {Lee}\ \emph {et~al.}(2021)\citenamefont {Lee},
  \citenamefont {Choe}, \citenamefont {Iaia}, \citenamefont {Li}, \citenamefont
  {Zhao}, \citenamefont {Shi}, \citenamefont {Ma}, \citenamefont {Yao},
  \citenamefont {Wang}, \citenamefont {Huang}, \citenamefont {Ochi},
  \citenamefont {Arita}, \citenamefont {Chatterjee}, \citenamefont {Morosan},
  \citenamefont {Madhavan},\ and\ \citenamefont {Trivedi}}]{Lee2021}%
  \BibitemOpen
  \bibfield  {author} {\bibinfo {author} {\bibfnamefont {Kyungmin}\
  \bibnamefont {Lee}}, \bibinfo {author} {\bibfnamefont {Jesse}\ \bibnamefont
  {Choe}}, \bibinfo {author} {\bibfnamefont {Davide}\ \bibnamefont {Iaia}},
  \bibinfo {author} {\bibfnamefont {Juqiang}\ \bibnamefont {Li}}, \bibinfo
  {author} {\bibfnamefont {Junjing}\ \bibnamefont {Zhao}}, \bibinfo {author}
  {\bibfnamefont {Ming}\ \bibnamefont {Shi}}, \bibinfo {author} {\bibfnamefont
  {Junzhang}\ \bibnamefont {Ma}}, \bibinfo {author} {\bibfnamefont {Mengyu}\
  \bibnamefont {Yao}}, \bibinfo {author} {\bibfnamefont {Zhenyu}\ \bibnamefont
  {Wang}}, \bibinfo {author} {\bibfnamefont {Chien-Lung}\ \bibnamefont
  {Huang}}, \bibinfo {author} {\bibfnamefont {Masayuki}\ \bibnamefont {Ochi}},
  \bibinfo {author} {\bibfnamefont {Ryotaro}\ \bibnamefont {Arita}}, \bibinfo
  {author} {\bibfnamefont {Utpal}\ \bibnamefont {Chatterjee}}, \bibinfo
  {author} {\bibfnamefont {Emilia}\ \bibnamefont {Morosan}}, \bibinfo {author}
  {\bibfnamefont {Vidya}\ \bibnamefont {Madhavan}}, \ and\ \bibinfo {author}
  {\bibfnamefont {Nandini}\ \bibnamefont {Trivedi}},\ }\bibfield  {title}
  {\enquote {\bibinfo {title} {Metal-to-insulator transition in pt-doped tise2
  driven by emergent network of narrow transport channels},}\ }\href {\doibase
  10.1038/s41535-020-00305-2} {\bibfield  {journal} {\bibinfo  {journal} {npj
  Quantum Materials}\ }\textbf {\bibinfo {volume} {6}},\ \bibinfo {pages} {8}
  (\bibinfo {year} {2021})}\BibitemShut {NoStop}%
\bibitem [{\citenamefont {Antonelli}\ \emph {et~al.}(2022)\citenamefont
  {Antonelli}, \citenamefont {Rahim}, \citenamefont {Watson}, \citenamefont
  {Rajan}, \citenamefont {Clark}, \citenamefont {Danilenko}, \citenamefont
  {Underwood}, \citenamefont {Markovi{\'{c}}}, \citenamefont {Abarca-Morales},
  \citenamefont {Kavanagh}, \citenamefont {Le~F{\`e}vre}, \citenamefont
  {Bertran}, \citenamefont {Rossnagel}, \citenamefont {Scanlon},\ and\
  \citenamefont {King}}]{Antonelli2022}%
  \BibitemOpen
  \bibfield  {author} {\bibinfo {author} {\bibfnamefont {Tommaso}\ \bibnamefont
  {Antonelli}}, \bibinfo {author} {\bibfnamefont {Warda}\ \bibnamefont
  {Rahim}}, \bibinfo {author} {\bibfnamefont {Matthew~D.}\ \bibnamefont
  {Watson}}, \bibinfo {author} {\bibfnamefont {Akhil}\ \bibnamefont {Rajan}},
  \bibinfo {author} {\bibfnamefont {Oliver~J.}\ \bibnamefont {Clark}}, \bibinfo
  {author} {\bibfnamefont {Alisa}\ \bibnamefont {Danilenko}}, \bibinfo {author}
  {\bibfnamefont {Kaycee}\ \bibnamefont {Underwood}}, \bibinfo {author}
  {\bibfnamefont {Igor}\ \bibnamefont {Markovi{\'{c}}}}, \bibinfo {author}
  {\bibfnamefont {Edgar}\ \bibnamefont {Abarca-Morales}}, \bibinfo {author}
  {\bibfnamefont {Se{\'a}n~R.}\ \bibnamefont {Kavanagh}}, \bibinfo {author}
  {\bibfnamefont {P.}~\bibnamefont {Le~F{\`e}vre}}, \bibinfo {author}
  {\bibfnamefont {F.}~\bibnamefont {Bertran}}, \bibinfo {author} {\bibfnamefont
  {K.}~\bibnamefont {Rossnagel}}, \bibinfo {author} {\bibfnamefont {David~O.}\
  \bibnamefont {Scanlon}}, \ and\ \bibinfo {author} {\bibfnamefont {Phil
  D.~C.}\ \bibnamefont {King}},\ }\bibfield  {title} {\enquote {\bibinfo
  {title} {Orbital-selective band hybridisation at the charge density wave
  transition in monolayer tite2},}\ }\href {\doibase
  10.1038/s41535-022-00508-9} {\bibfield  {journal} {\bibinfo  {journal} {npj
  Quantum Materials}\ }\textbf {\bibinfo {volume} {7}},\ \bibinfo {pages} {98}
  (\bibinfo {year} {2022})}\BibitemShut {NoStop}%
\bibitem [{\citenamefont {Lahneman}\ \emph {et~al.}(2022)\citenamefont
  {Lahneman}, \citenamefont {Slusar}, \citenamefont {Beringer}, \citenamefont
  {Jiang}, \citenamefont {Kim}, \citenamefont {Kim},\ and\ \citenamefont
  {Qazilbash}}]{Lahneman2022}%
  \BibitemOpen
  \bibfield  {author} {\bibinfo {author} {\bibfnamefont {D.~J.}\ \bibnamefont
  {Lahneman}}, \bibinfo {author} {\bibfnamefont {Tetiana}\ \bibnamefont
  {Slusar}}, \bibinfo {author} {\bibfnamefont {D.~B.}\ \bibnamefont
  {Beringer}}, \bibinfo {author} {\bibfnamefont {Haoyue}\ \bibnamefont
  {Jiang}}, \bibinfo {author} {\bibfnamefont {Chang-Yong}\ \bibnamefont {Kim}},
  \bibinfo {author} {\bibfnamefont {Hyun-Tak}\ \bibnamefont {Kim}}, \ and\
  \bibinfo {author} {\bibfnamefont {M.~M.}\ \bibnamefont {Qazilbash}},\
  }\bibfield  {title} {\enquote {\bibinfo {title} {Insulator-to-metal
  transition in ultrathin rutile vo2/tio2(001)},}\ }\href {\doibase
  10.1038/s41535-022-00479-x} {\bibfield  {journal} {\bibinfo  {journal} {npj
  Quantum Materials}\ }\textbf {\bibinfo {volume} {7}},\ \bibinfo {pages} {72}
  (\bibinfo {year} {2022})}\BibitemShut {NoStop}%
\bibitem [{\citenamefont {Luo}\ \emph {et~al.}(2022)\citenamefont {Luo},
  \citenamefont {Zhao}, \citenamefont {Zhou}, \citenamefont {Yang},
  \citenamefont {Fang}, \citenamefont {Yang}, \citenamefont {Gao},
  \citenamefont {Zhou},\ and\ \citenamefont {Zheng}}]{Luo2022}%
  \BibitemOpen
  \bibfield  {author} {\bibinfo {author} {\bibfnamefont {J.}~\bibnamefont
  {Luo}}, \bibinfo {author} {\bibfnamefont {Z.}~\bibnamefont {Zhao}}, \bibinfo
  {author} {\bibfnamefont {Y.~Z.}\ \bibnamefont {Zhou}}, \bibinfo {author}
  {\bibfnamefont {J.}~\bibnamefont {Yang}}, \bibinfo {author} {\bibfnamefont
  {A.~F.}\ \bibnamefont {Fang}}, \bibinfo {author} {\bibfnamefont {H.~T.}\
  \bibnamefont {Yang}}, \bibinfo {author} {\bibfnamefont {H.~J.}\ \bibnamefont
  {Gao}}, \bibinfo {author} {\bibfnamefont {R.}~\bibnamefont {Zhou}}, \ and\
  \bibinfo {author} {\bibfnamefont {Guo-qing}\ \bibnamefont {Zheng}},\
  }\bibfield  {title} {\enquote {\bibinfo {title} {Possible star-of-david
  pattern charge density wave with additional modulation in the kagome
  superconductor csv3sb5},}\ }\href {\doibase 10.1038/s41535-022-00437-7}
  {\bibfield  {journal} {\bibinfo  {journal} {npj Quantum Materials}\ }\textbf
  {\bibinfo {volume} {7}},\ \bibinfo {pages} {30} (\bibinfo {year}
  {2022})}\BibitemShut {NoStop}%
\bibitem [{\citenamefont {Profeta}\ and\ \citenamefont
  {Tosatti}(2007{\natexlab{a}})}]{PhysRevLett.98.086401}%
  \BibitemOpen
  \bibfield  {author} {\bibinfo {author} {\bibfnamefont {G.}~\bibnamefont
  {Profeta}}\ and\ \bibinfo {author} {\bibfnamefont {E.}~\bibnamefont
  {Tosatti}},\ }\bibfield  {title} {\enquote {\bibinfo {title} {{Triangular
  Mott-Hubbard Insulator Phases of $\mathrm{Sn}/\mathrm{Si}(111)$ and
  $\mathrm{Sn}/\mathrm{Ge}(111)$ Surfaces}},}\ }\href {\doibase
  10.1103/PhysRevLett.98.086401} {\bibfield  {journal} {\bibinfo  {journal}
  {Phys. Rev. Lett.}\ }\textbf {\bibinfo {volume} {98}},\ \bibinfo {pages}
  {086401} (\bibinfo {year} {2007}{\natexlab{a}})}\BibitemShut {NoStop}%
\bibitem [{\citenamefont {Hansmann}\ \emph {et~al.}(2016)\citenamefont
  {Hansmann}, \citenamefont {Ayral}, \citenamefont {Tejeda},\ and\
  \citenamefont {Biermann}}]{Hansmann2016}%
  \BibitemOpen
  \bibfield  {author} {\bibinfo {author} {\bibfnamefont {P.}~\bibnamefont
  {Hansmann}}, \bibinfo {author} {\bibfnamefont {T.}~\bibnamefont {Ayral}},
  \bibinfo {author} {\bibfnamefont {A.}~\bibnamefont {Tejeda}}, \ and\ \bibinfo
  {author} {\bibfnamefont {S.}~\bibnamefont {Biermann}},\ }\bibfield  {title}
  {\enquote {\bibinfo {title} {{Uncertainty principle for experimental
  measurements: Fast versus slow probes}},}\ }\href {\doibase
  10.1038/srep19728} {\bibfield  {journal} {\bibinfo  {journal} {Sci. Rep.}\
  }\textbf {\bibinfo {volume} {6}},\ \bibinfo {pages} {19728} (\bibinfo {year}
  {2016})}\BibitemShut {NoStop}%
\bibitem [{\citenamefont {Schuwalow}\ \emph {et~al.}(2010)\citenamefont
  {Schuwalow}, \citenamefont {Grieger},\ and\ \citenamefont
  {Lechermann}}]{PhysRevB.82.035116}%
  \BibitemOpen
  \bibfield  {author} {\bibinfo {author} {\bibfnamefont {Sergej}\ \bibnamefont
  {Schuwalow}}, \bibinfo {author} {\bibfnamefont {Daniel}\ \bibnamefont
  {Grieger}}, \ and\ \bibinfo {author} {\bibfnamefont {Frank}\ \bibnamefont
  {Lechermann}},\ }\bibfield  {title} {\enquote {\bibinfo {title} {{Realistic
  modeling of the electronic structure and the effect of correlations for
  Sn/Si(111) and Sn/Ge(111) surfaces}},}\ }\href {\doibase
  10.1103/PhysRevB.82.035116} {\bibfield  {journal} {\bibinfo  {journal} {Phys.
  Rev. B}\ }\textbf {\bibinfo {volume} {82}},\ \bibinfo {pages} {035116}
  (\bibinfo {year} {2010})}\BibitemShut {NoStop}%
\bibitem [{\citenamefont {Li}\ \emph {et~al.}(2011)\citenamefont {Li},
  \citenamefont {Laubach}, \citenamefont {Fleszar},\ and\ \citenamefont
  {Hanke}}]{PhysRevB.83.041104}%
  \BibitemOpen
  \bibfield  {author} {\bibinfo {author} {\bibfnamefont {Gang}\ \bibnamefont
  {Li}}, \bibinfo {author} {\bibfnamefont {Manuel}\ \bibnamefont {Laubach}},
  \bibinfo {author} {\bibfnamefont {Andrzej}\ \bibnamefont {Fleszar}}, \ and\
  \bibinfo {author} {\bibfnamefont {Werner}\ \bibnamefont {Hanke}},\ }\bibfield
   {title} {\enquote {\bibinfo {title} {{Geometrical frustration and the
  competing phases of the Sn/Si(111)
  $\sqrt{3}\ifmmode\times\else\texttimes\fi{}\sqrt{3}R{30}^{\ifmmode^\circ\else\textdegree\fi{}}$
  surface systems}},}\ }\href {\doibase 10.1103/PhysRevB.83.041104} {\bibfield
  {journal} {\bibinfo  {journal} {Phys. Rev. B}\ }\textbf {\bibinfo {volume}
  {83}},\ \bibinfo {pages} {041104} (\bibinfo {year} {2011})}\BibitemShut
  {NoStop}%
\bibitem [{\citenamefont {Adler}\ \emph {et~al.}(2019)\citenamefont {Adler},
  \citenamefont {Rachel}, \citenamefont {Laubach}, \citenamefont {Maklar},
  \citenamefont {Fleszar}, \citenamefont {Sch\"afer},\ and\ \citenamefont
  {Claessen}}]{PhysRevLett.123.086401}%
  \BibitemOpen
  \bibfield  {author} {\bibinfo {author} {\bibfnamefont {F.}~\bibnamefont
  {Adler}}, \bibinfo {author} {\bibfnamefont {S.}~\bibnamefont {Rachel}},
  \bibinfo {author} {\bibfnamefont {M.}~\bibnamefont {Laubach}}, \bibinfo
  {author} {\bibfnamefont {J.}~\bibnamefont {Maklar}}, \bibinfo {author}
  {\bibfnamefont {A.}~\bibnamefont {Fleszar}}, \bibinfo {author} {\bibfnamefont
  {J.}~\bibnamefont {Sch\"afer}}, \ and\ \bibinfo {author} {\bibfnamefont
  {R.}~\bibnamefont {Claessen}},\ }\bibfield  {title} {\enquote {\bibinfo
  {title} {Correlation-driven charge order in a frustrated two-dimensional atom
  lattice},}\ }\href {\doibase 10.1103/PhysRevLett.123.086401} {\bibfield
  {journal} {\bibinfo  {journal} {Phys. Rev. Lett.}\ }\textbf {\bibinfo
  {volume} {123}},\ \bibinfo {pages} {086401} (\bibinfo {year}
  {2019})}\BibitemShut {NoStop}%
\bibitem [{\citenamefont {Brun}\ \emph {et~al.}(2014)\citenamefont {Brun},
  \citenamefont {Cren}, \citenamefont {Cherkez}, \citenamefont {Debontridder},
  \citenamefont {Pons}, \citenamefont {Fokin}, \citenamefont {Tringides},
  \citenamefont {Bozhko}, \citenamefont {Ioffe}, \citenamefont {Altshuler},\
  and\ \citenamefont {Roditchev}}]{Brun2014}%
  \BibitemOpen
  \bibfield  {author} {\bibinfo {author} {\bibfnamefont {C.}~\bibnamefont
  {Brun}}, \bibinfo {author} {\bibfnamefont {T.}~\bibnamefont {Cren}}, \bibinfo
  {author} {\bibfnamefont {V.}~\bibnamefont {Cherkez}}, \bibinfo {author}
  {\bibfnamefont {F.}~\bibnamefont {Debontridder}}, \bibinfo {author}
  {\bibfnamefont {S.}~\bibnamefont {Pons}}, \bibinfo {author} {\bibfnamefont
  {D.}~\bibnamefont {Fokin}}, \bibinfo {author} {\bibfnamefont {M.~C.}\
  \bibnamefont {Tringides}}, \bibinfo {author} {\bibfnamefont {S.}~\bibnamefont
  {Bozhko}}, \bibinfo {author} {\bibfnamefont {L.~B.}\ \bibnamefont {Ioffe}},
  \bibinfo {author} {\bibfnamefont {B.~L.}\ \bibnamefont {Altshuler}}, \ and\
  \bibinfo {author} {\bibfnamefont {D.}~\bibnamefont {Roditchev}},\ }\bibfield
  {title} {\enquote {\bibinfo {title} {Remarkable effects of disorder on
  superconductivity of single atomic layers of lead on silicon},}\ }\href
  {\doibase 10.1038/nphys2937} {\bibfield  {journal} {\bibinfo  {journal} {Nat.
  Phys.}\ }\textbf {\bibinfo {volume} {10}},\ \bibinfo {pages} {444--450}
  (\bibinfo {year} {2014})}\BibitemShut {NoStop}%
\bibitem [{\citenamefont {Brun}\ \emph {et~al.}(2016)\citenamefont {Brun},
  \citenamefont {Cren},\ and\ \citenamefont {Roditchev}}]{Brun_2016}%
  \BibitemOpen
  \bibfield  {author} {\bibinfo {author} {\bibfnamefont {Christophe}\
  \bibnamefont {Brun}}, \bibinfo {author} {\bibfnamefont {Tristan}\
  \bibnamefont {Cren}}, \ and\ \bibinfo {author} {\bibfnamefont {Dimitri}\
  \bibnamefont {Roditchev}},\ }\bibfield  {title} {\enquote {\bibinfo {title}
  {{Review of 2D superconductivity: the ultimate case of epitaxial
  monolayers}},}\ }\href {\doibase 10.1088/0953-2048/30/1/013003} {\bibfield
  {journal} {\bibinfo  {journal} {Supercond. Sci. Technol.}\ }\textbf {\bibinfo
  {volume} {30}},\ \bibinfo {pages} {013003} (\bibinfo {year}
  {2016})}\BibitemShut {NoStop}%
\bibitem [{\citenamefont {Ming}\ \emph {et~al.}(2023)\citenamefont {Ming},
  \citenamefont {Wu}, \citenamefont {Chen}, \citenamefont {Wang}, \citenamefont
  {Mai}, \citenamefont {Maier}, \citenamefont {Strockoz}, \citenamefont
  {Venderbos}, \citenamefont {Gonz{\'a}lez}, \citenamefont {Ortega},
  \citenamefont {Johnston},\ and\ \citenamefont {Weitering}}]{Ming2023}%
  \BibitemOpen
  \bibfield  {author} {\bibinfo {author} {\bibfnamefont {F.}~\bibnamefont
  {Ming}}, \bibinfo {author} {\bibfnamefont {X.}~\bibnamefont {Wu}}, \bibinfo
  {author} {\bibfnamefont {C.}~\bibnamefont {Chen}}, \bibinfo {author}
  {\bibfnamefont {K.~D.}\ \bibnamefont {Wang}}, \bibinfo {author}
  {\bibfnamefont {P.}~\bibnamefont {Mai}}, \bibinfo {author} {\bibfnamefont
  {T.~A.}\ \bibnamefont {Maier}}, \bibinfo {author} {\bibfnamefont
  {J.}~\bibnamefont {Strockoz}}, \bibinfo {author} {\bibfnamefont {J.~W.~F.}\
  \bibnamefont {Venderbos}}, \bibinfo {author} {\bibfnamefont {C.}~\bibnamefont
  {Gonz{\'a}lez}}, \bibinfo {author} {\bibfnamefont {J.}~\bibnamefont
  {Ortega}}, \bibinfo {author} {\bibfnamefont {S.}~\bibnamefont {Johnston}}, \
  and\ \bibinfo {author} {\bibfnamefont {H.~H.}\ \bibnamefont {Weitering}},\
  }\bibfield  {title} {\enquote {\bibinfo {title} {Evidence for chiral
  superconductivity on a silicon surface},}\ }\href {\doibase
  10.1038/s41567-022-01889-1} {\bibfield  {journal} {\bibinfo  {journal}
  {Nature Physics}\ }\textbf {\bibinfo {volume} {19}},\ \bibinfo {pages}
  {500--506} (\bibinfo {year} {2023})}\BibitemShut {NoStop}%
\bibitem [{\citenamefont {Gor'kov}\ and\ \citenamefont
  {Rashba}(2001)}]{PhysRevLett.87.037004}%
  \BibitemOpen
  \bibfield  {author} {\bibinfo {author} {\bibfnamefont {Lev~P.}\ \bibnamefont
  {Gor'kov}}\ and\ \bibinfo {author} {\bibfnamefont {Emmanuel~I.}\ \bibnamefont
  {Rashba}},\ }\bibfield  {title} {\enquote {\bibinfo {title} {{Superconducting
  2D System with Lifted Spin Degeneracy: Mixed Singlet-Triplet State}},}\
  }\href {\doibase 10.1103/PhysRevLett.87.037004} {\bibfield  {journal}
  {\bibinfo  {journal} {Phys. Rev. Lett.}\ }\textbf {\bibinfo {volume} {87}},\
  \bibinfo {pages} {037004} (\bibinfo {year} {2001})}\BibitemShut {NoStop}%
\bibitem [{\citenamefont {Brihuega}\ \emph {et~al.}(2005)\citenamefont
  {Brihuega}, \citenamefont {Custance}, \citenamefont {P\'erez},\ and\
  \citenamefont {G\'omez-Rodriguez}}]{PhysRevLett.94.046101}%
  \BibitemOpen
  \bibfield  {author} {\bibinfo {author} {\bibfnamefont {I.}~\bibnamefont
  {Brihuega}}, \bibinfo {author} {\bibfnamefont {O.}~\bibnamefont {Custance}},
  \bibinfo {author} {\bibfnamefont {Rub\'en}\ \bibnamefont {P\'erez}}, \ and\
  \bibinfo {author} {\bibfnamefont {J.~M.}\ \bibnamefont {G\'omez-Rodriguez}},\
  }\bibfield  {title} {\enquote {\bibinfo {title} {Intrinsic character of the
  $(3\ifmmode\times\else\texttimes\fi{}3)$ to
  $(\sqrt{3}\ifmmode\times\else\texttimes\fi{}\sqrt{3})$ phase transition in
  $\mathrm{P}\mathrm{b}/\mathrm{S}\mathrm{i}(111)$},}\ }\href {\doibase
  10.1103/PhysRevLett.94.046101} {\bibfield  {journal} {\bibinfo  {journal}
  {Phys. Rev. Lett.}\ }\textbf {\bibinfo {volume} {94}},\ \bibinfo {pages}
  {046101} (\bibinfo {year} {2005})}\BibitemShut {NoStop}%
\bibitem [{\citenamefont {Brihuega}\ \emph {et~al.}(2007)\citenamefont
  {Brihuega}, \citenamefont {Custance}, \citenamefont {Ugeda},\ and\
  \citenamefont {G\'omez-Rodriguez}}]{PhysRevB.75.155411}%
  \BibitemOpen
  \bibfield  {author} {\bibinfo {author} {\bibfnamefont {I.}~\bibnamefont
  {Brihuega}}, \bibinfo {author} {\bibfnamefont {O.}~\bibnamefont {Custance}},
  \bibinfo {author} {\bibfnamefont {M.~M.}\ \bibnamefont {Ugeda}}, \ and\
  \bibinfo {author} {\bibfnamefont {J.~M.}\ \bibnamefont {G\'omez-Rodriguez}},\
  }\bibfield  {title} {\enquote {\bibinfo {title} {Defects in the
  $(\sqrt{3}\ifmmode\times\else\texttimes\fi{}\sqrt{3})\ensuremath{\Leftrightarrow}(3\ifmmode\times\else\texttimes\fi{}3)$
  phase transition in $\mathrm{P}\mathrm{b}/\mathrm{S}\mathrm{i}(111)$
  system},}\ }\href {\doibase 10.1103/PhysRevB.75.155411} {\bibfield  {journal}
  {\bibinfo  {journal} {Phys. Rev. B}\ }\textbf {\bibinfo {volume} {75}},\
  \bibinfo {pages} {155411} (\bibinfo {year} {2007})}\BibitemShut {NoStop}%
\bibitem [{\citenamefont {{Tresca, C. and Bilgeri, T. and M\'enard, G. and
  Cherkez, V. and Federicci, R. and Longo, D. and Herv\'e, M. and Debontridder,
  F. and David, P. and Roditchev, D. and Profeta, G. and Cren, T. and Calandra,
  M. and Brun, C.}}(2023)}]{PhysRevB.107.035125}%
  \BibitemOpen
  \bibfield  {author} {\bibinfo {author} {\bibnamefont {{Tresca, C. and
  Bilgeri, T. and M\'enard, G. and Cherkez, V. and Federicci, R. and Longo, D.
  and Herv\'e, M. and Debontridder, F. and David, P. and Roditchev, D. and
  Profeta, G. and Cren, T. and Calandra, M. and Brun, C.}}},\ }\bibfield
  {title} {\enquote {\bibinfo {title} {{Importance of accurately measuring LDOS
  maps using scanning tunneling spectroscopy in materials presenting
  atom-dependent charge order: The case of the correlated Pb/Si(111) single
  atomic layer}},}\ }\href {\doibase 10.1103/PhysRevB.107.035125} {\bibfield
  {journal} {\bibinfo  {journal} {Phys. Rev. B}\ }\textbf {\bibinfo {volume}
  {107}},\ \bibinfo {pages} {035125} (\bibinfo {year} {2023})}\BibitemShut
  {NoStop}%
\bibitem [{\citenamefont {Carpinelli}\ \emph {et~al.}(1996)\citenamefont
  {Carpinelli}, \citenamefont {Weitering}, \citenamefont {Plummer},\ and\
  \citenamefont {Stumpf}}]{Carpinelli1996}%
  \BibitemOpen
  \bibfield  {author} {\bibinfo {author} {\bibfnamefont {Joseph~M.}\
  \bibnamefont {Carpinelli}}, \bibinfo {author} {\bibfnamefont {Hanno~H.}\
  \bibnamefont {Weitering}}, \bibinfo {author} {\bibfnamefont {E.~Ward}\
  \bibnamefont {Plummer}}, \ and\ \bibinfo {author} {\bibfnamefont {Roland}\
  \bibnamefont {Stumpf}},\ }\bibfield  {title} {\enquote {\bibinfo {title}
  {Direct observation of a surface charge density wave},}\ }\href {\doibase
  10.1038/381398a0} {\bibfield  {journal} {\bibinfo  {journal} {Nature}\
  }\textbf {\bibinfo {volume} {381}},\ \bibinfo {pages} {398--400} (\bibinfo
  {year} {1996})}\BibitemShut {NoStop}%
\bibitem [{\citenamefont {Ming}\ \emph {et~al.}(2017)\citenamefont {Ming},
  \citenamefont {Johnston}, \citenamefont {Mulugeta}, \citenamefont {Smith},
  \citenamefont {Vilmercati}, \citenamefont {Lee}, \citenamefont {Maier},
  \citenamefont {Snijders},\ and\ \citenamefont
  {Weitering}}]{PhysRevLett.119.266802}%
  \BibitemOpen
  \bibfield  {author} {\bibinfo {author} {\bibfnamefont {Fangfei}\ \bibnamefont
  {Ming}}, \bibinfo {author} {\bibfnamefont {Steve}\ \bibnamefont {Johnston}},
  \bibinfo {author} {\bibfnamefont {Daniel}\ \bibnamefont {Mulugeta}}, \bibinfo
  {author} {\bibfnamefont {Tyler~S.}\ \bibnamefont {Smith}}, \bibinfo {author}
  {\bibfnamefont {Paolo}\ \bibnamefont {Vilmercati}}, \bibinfo {author}
  {\bibfnamefont {Geunseop}\ \bibnamefont {Lee}}, \bibinfo {author}
  {\bibfnamefont {Thomas~A.}\ \bibnamefont {Maier}}, \bibinfo {author}
  {\bibfnamefont {Paul~C.}\ \bibnamefont {Snijders}}, \ and\ \bibinfo {author}
  {\bibfnamefont {Hanno~H.}\ \bibnamefont {Weitering}},\ }\bibfield  {title}
  {\enquote {\bibinfo {title} {{Realization of a Hole-Doped Mott Insulator on a
  Triangular Silicon Lattice}},}\ }\href {\doibase
  10.1103/PhysRevLett.119.266802} {\bibfield  {journal} {\bibinfo  {journal}
  {Phys. Rev. Lett.}\ }\textbf {\bibinfo {volume} {119}},\ \bibinfo {pages}
  {266802} (\bibinfo {year} {2017})}\BibitemShut {NoStop}%
\bibitem [{\citenamefont {Wu}\ \emph {et~al.}(2020)\citenamefont {Wu},
  \citenamefont {Ming}, \citenamefont {Smith}, \citenamefont {Liu},
  \citenamefont {Ye}, \citenamefont {Wang}, \citenamefont {Johnston},\ and\
  \citenamefont {Weitering}}]{PhysRevLett.125.117001}%
  \BibitemOpen
  \bibfield  {author} {\bibinfo {author} {\bibfnamefont {Xuefeng}\ \bibnamefont
  {Wu}}, \bibinfo {author} {\bibfnamefont {Fangfei}\ \bibnamefont {Ming}},
  \bibinfo {author} {\bibfnamefont {Tyler~S.}\ \bibnamefont {Smith}}, \bibinfo
  {author} {\bibfnamefont {Guowei}\ \bibnamefont {Liu}}, \bibinfo {author}
  {\bibfnamefont {Fei}\ \bibnamefont {Ye}}, \bibinfo {author} {\bibfnamefont
  {Kedong}\ \bibnamefont {Wang}}, \bibinfo {author} {\bibfnamefont {Steven}\
  \bibnamefont {Johnston}}, \ and\ \bibinfo {author} {\bibfnamefont {Hanno~H.}\
  \bibnamefont {Weitering}},\ }\bibfield  {title} {\enquote {\bibinfo {title}
  {{Superconductivity in a Hole-Doped Mott-Insulating Triangular Adatom Layer
  on a Silicon Surface}},}\ }\href {\doibase 10.1103/PhysRevLett.125.117001}
  {\bibfield  {journal} {\bibinfo  {journal} {Phys. Rev. Lett.}\ }\textbf
  {\bibinfo {volume} {125}},\ \bibinfo {pages} {117001} (\bibinfo {year}
  {2020})}\BibitemShut {NoStop}%
\bibitem [{\citenamefont {Tresca~et al.}(2018)}]{PhysRevLett.120.196402_sm}%
  \BibitemOpen
  \bibfield  {author} {\bibinfo {author} {\bibfnamefont {C.}~\bibnamefont
  {Tresca~et al.}},\ }\bibfield  {title} {\enquote {\bibinfo {title} {{Chiral
  Spin Texture in the Charge-Density-Wave Phase of the Correlated Metallic
  $\mathrm{Pb}/\mathrm{Si}(111)$ Monolayer}},}\ }\href {\doibase
  10.1103/PhysRevLett.120.196402} {\bibfield  {journal} {\bibinfo  {journal}
  {Phys. Rev. Lett.}\ }\textbf {\bibinfo {volume} {120}},\ \bibinfo {pages}
  {196402} (\bibinfo {year} {2018})},\ \bibinfo {note} {{Supplemental
  Material}}\BibitemShut {NoStop}%
\bibitem [{\citenamefont {Adler~et al.}(2019)}]{PhysRevLett.123.086401_sm}%
  \BibitemOpen
  \bibfield  {author} {\bibinfo {author} {\bibfnamefont {F.}~\bibnamefont
  {Adler~et al.}},\ }\bibfield  {title} {\enquote {\bibinfo {title}
  {{Correlation-Driven Charge Order in a Frustrated Two-Dimensional Atom
  Lattice}},}\ }\href {\doibase 10.1103/PhysRevLett.123.086401} {\bibfield
  {journal} {\bibinfo  {journal} {Phys. Rev. Lett.}\ }\textbf {\bibinfo
  {volume} {123}},\ \bibinfo {pages} {086401} (\bibinfo {year} {2019})},\
  \bibinfo {note} {{Supplemental Material}}\BibitemShut {NoStop}%
\bibitem [{\citenamefont {Bychkov}\ and\ \citenamefont
  {Rashba}(1984)}]{Rashba}%
  \BibitemOpen
  \bibfield  {author} {\bibinfo {author} {\bibfnamefont {Y.~A.}\ \bibnamefont
  {Bychkov}}\ and\ \bibinfo {author} {\bibfnamefont {E.~I.}\ \bibnamefont
  {Rashba}},\ }\bibfield  {title} {\enquote {\bibinfo {title} {{Properties of a
  2D electron gas with lifted spectral degeneracy}},}\ }\href@noop {}
  {\bibfield  {journal} {\bibinfo  {journal} {Pis'ma Zh. Eksp. Teor. Fiz.}\
  }\textbf {\bibinfo {volume} {39}},\ \bibinfo {pages} {66} (\bibinfo {year}
  {1984})},\ \bibinfo {note} {[JETP Lett. {\bf 39}, 78 (1984)]}\BibitemShut
  {NoStop}%
\bibitem [{\citenamefont {Yildirim}\ \emph {et~al.}(1995)\citenamefont
  {Yildirim}, \citenamefont {Harris}, \citenamefont {Aharony},\ and\
  \citenamefont {Entin-Wohlman}}]{PhysRevB.52.10239}%
  \BibitemOpen
  \bibfield  {author} {\bibinfo {author} {\bibfnamefont {T.}~\bibnamefont
  {Yildirim}}, \bibinfo {author} {\bibfnamefont {A.~B.}\ \bibnamefont
  {Harris}}, \bibinfo {author} {\bibfnamefont {Amnon}\ \bibnamefont {Aharony}},
  \ and\ \bibinfo {author} {\bibfnamefont {O.}~\bibnamefont {Entin-Wohlman}},\
  }\bibfield  {title} {\enquote {\bibinfo {title} {{Anisotropic spin
  Hamiltonians due to spin-orbit and Coulomb exchange interactions}},}\ }\href
  {\doibase 10.1103/PhysRevB.52.10239} {\bibfield  {journal} {\bibinfo
  {journal} {Phys. Rev. B}\ }\textbf {\bibinfo {volume} {52}},\ \bibinfo
  {pages} {10239} (\bibinfo {year} {1995})}\BibitemShut {NoStop}%
\bibitem [{\citenamefont {Avila}\ \emph {et~al.}(1999)\citenamefont {Avila},
  \citenamefont {Mascaraque}, \citenamefont {Michel}, \citenamefont {Asensio},
  \citenamefont {LeLay}, \citenamefont {Ortega}, \citenamefont {P\'erez},\ and\
  \citenamefont {Flores}}]{PhysRevLett.82.442}%
  \BibitemOpen
  \bibfield  {author} {\bibinfo {author} {\bibfnamefont {J.}~\bibnamefont
  {Avila}}, \bibinfo {author} {\bibfnamefont {A.}~\bibnamefont {Mascaraque}},
  \bibinfo {author} {\bibfnamefont {E.~G.}\ \bibnamefont {Michel}}, \bibinfo
  {author} {\bibfnamefont {M.~C.}\ \bibnamefont {Asensio}}, \bibinfo {author}
  {\bibfnamefont {G.}~\bibnamefont {LeLay}}, \bibinfo {author} {\bibfnamefont
  {J.}~\bibnamefont {Ortega}}, \bibinfo {author} {\bibfnamefont
  {R.}~\bibnamefont {P\'erez}}, \ and\ \bibinfo {author} {\bibfnamefont
  {F.}~\bibnamefont {Flores}},\ }\bibfield  {title} {\enquote {\bibinfo {title}
  {{Dynamical Fluctuations as the Origin of a Surface Phase Transition in
  $\mathrm{Sn}/\mathrm{Ge}(111)$}},}\ }\href {\doibase
  10.1103/PhysRevLett.82.442} {\bibfield  {journal} {\bibinfo  {journal} {Phys.
  Rev. Lett.}\ }\textbf {\bibinfo {volume} {82}},\ \bibinfo {pages} {442--445}
  (\bibinfo {year} {1999})}\BibitemShut {NoStop}%
\bibitem [{\citenamefont {Dudr}\ \emph {et~al.}(2004)\citenamefont {Dudr},
  \citenamefont {Tsud}, \citenamefont {Fab\'{\i}k}, \citenamefont
  {Vondr\'a\ifmmode~\check{c}\else \v{c}\fi{}ek}, \citenamefont {Matol\'{\i}n},
  \citenamefont {Ch\'ab},\ and\ \citenamefont {Prince}}]{PhysRevB.70.155334}%
  \BibitemOpen
  \bibfield  {author} {\bibinfo {author} {\bibfnamefont {Viktor}\ \bibnamefont
  {Dudr}}, \bibinfo {author} {\bibfnamefont {Natalia}\ \bibnamefont {Tsud}},
  \bibinfo {author} {\bibfnamefont {Stanislav}\ \bibnamefont {Fab\'{\i}k}},
  \bibinfo {author} {\bibfnamefont {Martin}\ \bibnamefont
  {Vondr\'a\ifmmode~\check{c}\else \v{c}\fi{}ek}}, \bibinfo {author}
  {\bibfnamefont {Vladim\'{\i}r}\ \bibnamefont {Matol\'{\i}n}}, \bibinfo
  {author} {\bibfnamefont {Vladim\'{\i}r}\ \bibnamefont {Ch\'ab}}, \ and\
  \bibinfo {author} {\bibfnamefont {Kevin~C.}\ \bibnamefont {Prince}},\
  }\bibfield  {title} {\enquote {\bibinfo {title} {{Evidence for valence-charge
  fluctuations in the
  $\sqrt{3}\ifmmode\times\else\texttimes\fi{}\sqrt{3}\text{\ensuremath{-}}\mathrm{Pb}/\mathrm{Si}(111)$
  system}},}\ }\href {\doibase 10.1103/PhysRevB.70.155334} {\bibfield
  {journal} {\bibinfo  {journal} {Phys. Rev. B}\ }\textbf {\bibinfo {volume}
  {70}},\ \bibinfo {pages} {155334} (\bibinfo {year} {2004})}\BibitemShut
  {NoStop}%
\bibitem [{\citenamefont {Punk}\ \emph {et~al.}(2014)\citenamefont {Punk},
  \citenamefont {Chowdhury},\ and\ \citenamefont
  {Sachdev}}]{punk2014topological}%
  \BibitemOpen
  \bibfield  {author} {\bibinfo {author} {\bibfnamefont {Matthias}\
  \bibnamefont {Punk}}, \bibinfo {author} {\bibfnamefont {Debanjan}\
  \bibnamefont {Chowdhury}}, \ and\ \bibinfo {author} {\bibfnamefont {Subir}\
  \bibnamefont {Sachdev}},\ }\bibfield  {title} {\enquote {\bibinfo {title}
  {{Topological excitations and the dynamic structure factor of spin liquids on
  the kagome lattice}},}\ }\href {\doibase 10.1038/nphys2887} {\bibfield
  {journal} {\bibinfo  {journal} {Nat. Phys.}\ }\textbf {\bibinfo {volume}
  {10}},\ \bibinfo {pages} {289--293} (\bibinfo {year} {2014})}\BibitemShut
  {NoStop}%
\bibitem [{\citenamefont {Sherman}\ and\ \citenamefont
  {Singh}(2018)}]{PhysRevB.97.014423}%
  \BibitemOpen
  \bibfield  {author} {\bibinfo {author} {\bibfnamefont {Nicholas~E.}\
  \bibnamefont {Sherman}}\ and\ \bibinfo {author} {\bibfnamefont {Rajiv R.~P.}\
  \bibnamefont {Singh}},\ }\bibfield  {title} {\enquote {\bibinfo {title}
  {{Structure factors of the kagome-lattice Heisenberg antiferromagnets at
  finite temperatures}},}\ }\href {\doibase 10.1103/PhysRevB.97.014423}
  {\bibfield  {journal} {\bibinfo  {journal} {Phys. Rev. B}\ }\textbf {\bibinfo
  {volume} {97}},\ \bibinfo {pages} {014423} (\bibinfo {year}
  {2018})}\BibitemShut {NoStop}%
\bibitem [{\citenamefont {Kaufmann}\ \emph {et~al.}(2021)\citenamefont
  {Kaufmann}, \citenamefont {Steiner}, \citenamefont {Scalettar}, \citenamefont
  {Held},\ and\ \citenamefont {Janson}}]{PhysRevB.104.165127}%
  \BibitemOpen
  \bibfield  {author} {\bibinfo {author} {\bibfnamefont {Josef}\ \bibnamefont
  {Kaufmann}}, \bibinfo {author} {\bibfnamefont {Klaus}\ \bibnamefont
  {Steiner}}, \bibinfo {author} {\bibfnamefont {Richard~T.}\ \bibnamefont
  {Scalettar}}, \bibinfo {author} {\bibfnamefont {Karsten}\ \bibnamefont
  {Held}}, \ and\ \bibinfo {author} {\bibfnamefont {Oleg}\ \bibnamefont
  {Janson}},\ }\bibfield  {title} {\enquote {\bibinfo {title} {{How
  correlations change the magnetic structure factor of the kagome Hubbard
  model}},}\ }\href {\doibase 10.1103/PhysRevB.104.165127} {\bibfield
  {journal} {\bibinfo  {journal} {Phys. Rev. B}\ }\textbf {\bibinfo {volume}
  {104}},\ \bibinfo {pages} {165127} (\bibinfo {year} {2021})}\BibitemShut
  {NoStop}%
\bibitem [{\citenamefont {Stepanov}\ \emph
  {et~al.}(2019{\natexlab{a}})\citenamefont {Stepanov}, \citenamefont
  {Harkov},\ and\ \citenamefont {Lichtenstein}}]{PhysRevB.100.205115}%
  \BibitemOpen
  \bibfield  {author} {\bibinfo {author} {\bibfnamefont {E.~A.}\ \bibnamefont
  {Stepanov}}, \bibinfo {author} {\bibfnamefont {V.}~\bibnamefont {Harkov}}, \
  and\ \bibinfo {author} {\bibfnamefont {A.~I.}\ \bibnamefont {Lichtenstein}},\
  }\bibfield  {title} {\enquote {\bibinfo {title} {{Consistent partial
  bosonization of the extended Hubbard model}},}\ }\href {\doibase
  10.1103/PhysRevB.100.205115} {\bibfield  {journal} {\bibinfo  {journal}
  {Phys. Rev. B}\ }\textbf {\bibinfo {volume} {100}},\ \bibinfo {pages}
  {205115} (\bibinfo {year} {2019}{\natexlab{a}})}\BibitemShut {NoStop}%
\bibitem [{\citenamefont {Harkov}\ \emph {et~al.}(2021)\citenamefont {Harkov},
  \citenamefont {Vandelli}, \citenamefont {Brener}, \citenamefont
  {Lichtenstein},\ and\ \citenamefont {Stepanov}}]{PhysRevB.103.245123}%
  \BibitemOpen
  \bibfield  {author} {\bibinfo {author} {\bibfnamefont {V.}~\bibnamefont
  {Harkov}}, \bibinfo {author} {\bibfnamefont {M.}~\bibnamefont {Vandelli}},
  \bibinfo {author} {\bibfnamefont {S.}~\bibnamefont {Brener}}, \bibinfo
  {author} {\bibfnamefont {A.~I.}\ \bibnamefont {Lichtenstein}}, \ and\
  \bibinfo {author} {\bibfnamefont {E.~A.}\ \bibnamefont {Stepanov}},\
  }\bibfield  {title} {\enquote {\bibinfo {title} {Impact of partially
  bosonized collective fluctuations on electronic degrees of freedom},}\ }\href
  {\doibase 10.1103/PhysRevB.103.245123} {\bibfield  {journal} {\bibinfo
  {journal} {Phys. Rev. B}\ }\textbf {\bibinfo {volume} {103}},\ \bibinfo
  {pages} {245123} (\bibinfo {year} {2021})}\BibitemShut {NoStop}%
\bibitem [{\citenamefont {Vandelli}\ \emph {et~al.}(2022)\citenamefont
  {Vandelli}, \citenamefont {Kaufmann}, \citenamefont {El-Nabulsi},
  \citenamefont {Harkov}, \citenamefont {Lichtenstein},\ and\ \citenamefont
  {Stepanov}}]{10.21468/SciPostPhys.13.2.036}%
  \BibitemOpen
  \bibfield  {author} {\bibinfo {author} {\bibfnamefont {Matteo}\ \bibnamefont
  {Vandelli}}, \bibinfo {author} {\bibfnamefont {Josef}\ \bibnamefont
  {Kaufmann}}, \bibinfo {author} {\bibfnamefont {Mohammed}\ \bibnamefont
  {El-Nabulsi}}, \bibinfo {author} {\bibfnamefont {Viktor}\ \bibnamefont
  {Harkov}}, \bibinfo {author} {\bibfnamefont {Alexander~I.}\ \bibnamefont
  {Lichtenstein}}, \ and\ \bibinfo {author} {\bibfnamefont {Evgeny~A.}\
  \bibnamefont {Stepanov}},\ }\bibfield  {title} {\enquote {\bibinfo {title}
  {{Multi-band D-TRILEX approach to materials with strong electronic
  correlations}},}\ }\href {\doibase 10.21468/SciPostPhys.13.2.036} {\bibfield
  {journal} {\bibinfo  {journal} {SciPost Phys.}\ }\textbf {\bibinfo {volume}
  {13}},\ \bibinfo {pages} {036} (\bibinfo {year} {2022})}\BibitemShut
  {NoStop}%
\bibitem [{\citenamefont {Stepanov}\ \emph {et~al.}(2021)\citenamefont
  {Stepanov}, \citenamefont {Nomura}, \citenamefont {Lichtenstein},\ and\
  \citenamefont {Biermann}}]{PhysRevLett.127.207205}%
  \BibitemOpen
  \bibfield  {author} {\bibinfo {author} {\bibfnamefont {Evgeny~A.}\
  \bibnamefont {Stepanov}}, \bibinfo {author} {\bibfnamefont {Yusuke}\
  \bibnamefont {Nomura}}, \bibinfo {author} {\bibfnamefont {Alexander~I.}\
  \bibnamefont {Lichtenstein}}, \ and\ \bibinfo {author} {\bibfnamefont
  {Silke}\ \bibnamefont {Biermann}},\ }\bibfield  {title} {\enquote {\bibinfo
  {title} {{Orbital Isotropy of Magnetic Fluctuations in Correlated Electron
  Materials Induced by Hund's Exchange Coupling}},}\ }\href {\doibase
  10.1103/PhysRevLett.127.207205} {\bibfield  {journal} {\bibinfo  {journal}
  {Phys. Rev. Lett.}\ }\textbf {\bibinfo {volume} {127}},\ \bibinfo {pages}
  {207205} (\bibinfo {year} {2021})}\BibitemShut {NoStop}%
\bibitem [{\citenamefont {Stepanov}\ \emph
  {et~al.}(2022{\natexlab{a}})\citenamefont {Stepanov}, \citenamefont {Harkov},
  \citenamefont {R\"osner}, \citenamefont {Lichtenstein}, \citenamefont
  {Katsnelson},\ and\ \citenamefont {Rudenko}}]{stepanov2021coexisting}%
  \BibitemOpen
  \bibfield  {author} {\bibinfo {author} {\bibfnamefont {E.~A.}\ \bibnamefont
  {Stepanov}}, \bibinfo {author} {\bibfnamefont {V.}~\bibnamefont {Harkov}},
  \bibinfo {author} {\bibfnamefont {M.}~\bibnamefont {R\"osner}}, \bibinfo
  {author} {\bibfnamefont {A.~I.}\ \bibnamefont {Lichtenstein}}, \bibinfo
  {author} {\bibfnamefont {M.~I.}\ \bibnamefont {Katsnelson}}, \ and\ \bibinfo
  {author} {\bibfnamefont {A.~N.}\ \bibnamefont {Rudenko}},\ }\bibfield
  {title} {\enquote {\bibinfo {title} {{Coexisting charge density wave and
  ferromagnetic instabilities in monolayer InSe}},}\ }\href {\doibase
  10.1038/s41524-022-00798-4} {\bibfield  {journal} {\bibinfo  {journal} {npj
  Comput. Mater.}\ }\textbf {\bibinfo {volume} {8}},\ \bibinfo {pages} {118}
  (\bibinfo {year} {2022}{\natexlab{a}})}\BibitemShut {NoStop}%
\bibitem [{\citenamefont {Stepanov}(2022)}]{PhysRevLett.129.096404}%
  \BibitemOpen
  \bibfield  {author} {\bibinfo {author} {\bibfnamefont {Evgeny~A.}\
  \bibnamefont {Stepanov}},\ }\bibfield  {title} {\enquote {\bibinfo {title}
  {{Eliminating Orbital Selectivity from the Metal-Insulator Transition by
  Strong Magnetic Fluctuations}},}\ }\href {\doibase
  10.1103/PhysRevLett.129.096404} {\bibfield  {journal} {\bibinfo  {journal}
  {Phys. Rev. Lett.}\ }\textbf {\bibinfo {volume} {129}},\ \bibinfo {pages}
  {096404} (\bibinfo {year} {2022})}\BibitemShut {NoStop}%
\bibitem [{\citenamefont {{Vandelli}}\ \emph {et~al.}(2022)\citenamefont
  {{Vandelli}}, \citenamefont {{Kaufmann}}, \citenamefont {{Harkov}},
  \citenamefont {{Lichtenstein}}, \citenamefont {{Held}},\ and\ \citenamefont
  {{Stepanov}}}]{2022arXiv220402116V}%
  \BibitemOpen
  \bibfield  {author} {\bibinfo {author} {\bibfnamefont {M.}~\bibnamefont
  {{Vandelli}}}, \bibinfo {author} {\bibfnamefont {J.}~\bibnamefont
  {{Kaufmann}}}, \bibinfo {author} {\bibfnamefont {V.}~\bibnamefont
  {{Harkov}}}, \bibinfo {author} {\bibfnamefont {A.~I.}\ \bibnamefont
  {{Lichtenstein}}}, \bibinfo {author} {\bibfnamefont {K.}~\bibnamefont
  {{Held}}}, \ and\ \bibinfo {author} {\bibfnamefont {E.~A.}\ \bibnamefont
  {{Stepanov}}},\ }\href@noop {} {\enquote {\bibinfo {title} {{Extended regime
  of coexisting metallic and insulating phases in a two-orbital electronic
  system}},}\ }\bibinfo {howpublished} {Preprint at
  \url{https://arxiv.org/abs/2204.02116}} (\bibinfo {year} {2022})\BibitemShut
  {NoStop}%
\bibitem [{\citenamefont {Kurz}\ \emph {et~al.}(2001)\citenamefont {Kurz},
  \citenamefont {Bihlmayer}, \citenamefont {Hirai},\ and\ \citenamefont
  {Bl\"ugel}}]{PhysRevLett.86.1106}%
  \BibitemOpen
  \bibfield  {author} {\bibinfo {author} {\bibfnamefont {Ph.}\ \bibnamefont
  {Kurz}}, \bibinfo {author} {\bibfnamefont {G.}~\bibnamefont {Bihlmayer}},
  \bibinfo {author} {\bibfnamefont {K.}~\bibnamefont {Hirai}}, \ and\ \bibinfo
  {author} {\bibfnamefont {S.}~\bibnamefont {Bl\"ugel}},\ }\bibfield  {title}
  {\enquote {\bibinfo {title} {{Three-Dimensional Spin Structure on a
  Two-Dimensional Lattice: Mn/Cu(111)}},}\ }\href {\doibase
  10.1103/PhysRevLett.86.1106} {\bibfield  {journal} {\bibinfo  {journal}
  {Phys. Rev. Lett.}\ }\textbf {\bibinfo {volume} {86}},\ \bibinfo {pages}
  {1106--1109} (\bibinfo {year} {2001})}\BibitemShut {NoStop}%
\bibitem [{\citenamefont {Cort\'es}\ \emph {et~al.}(2013)\citenamefont
  {Cort\'es}, \citenamefont {Tejeda}, \citenamefont {Lobo-Checa}, \citenamefont
  {Didiot}, \citenamefont {Kierren}, \citenamefont {Malterre}, \citenamefont
  {Merino}, \citenamefont {Flores}, \citenamefont {Michel},\ and\ \citenamefont
  {Mascaraque}}]{PhysRevB.88.125113}%
  \BibitemOpen
  \bibfield  {author} {\bibinfo {author} {\bibfnamefont {R.}~\bibnamefont
  {Cort\'es}}, \bibinfo {author} {\bibfnamefont {A.}~\bibnamefont {Tejeda}},
  \bibinfo {author} {\bibfnamefont {J.}~\bibnamefont {Lobo-Checa}}, \bibinfo
  {author} {\bibfnamefont {C.}~\bibnamefont {Didiot}}, \bibinfo {author}
  {\bibfnamefont {B.}~\bibnamefont {Kierren}}, \bibinfo {author} {\bibfnamefont
  {D.}~\bibnamefont {Malterre}}, \bibinfo {author} {\bibfnamefont
  {J.}~\bibnamefont {Merino}}, \bibinfo {author} {\bibfnamefont
  {F.}~\bibnamefont {Flores}}, \bibinfo {author} {\bibfnamefont {E.~G.}\
  \bibnamefont {Michel}}, \ and\ \bibinfo {author} {\bibfnamefont
  {A.}~\bibnamefont {Mascaraque}},\ }\bibfield  {title} {\enquote {\bibinfo
  {title} {{Competing charge ordering and Mott phases in a correlated
  Sn/Ge(111) two-dimensional triangular lattice}},}\ }\href {\doibase
  10.1103/PhysRevB.88.125113} {\bibfield  {journal} {\bibinfo  {journal} {Phys.
  Rev. B}\ }\textbf {\bibinfo {volume} {88}},\ \bibinfo {pages} {125113}
  (\bibinfo {year} {2013})}\BibitemShut {NoStop}%
\bibitem [{\citenamefont {Ressel}\ \emph {et~al.}(2002)\citenamefont {Ressel},
  \citenamefont {Slez\'ak}, \citenamefont {Prince},\ and\ \citenamefont
  {Ch\'ab}}]{PhysRevB.66.035325}%
  \BibitemOpen
  \bibfield  {author} {\bibinfo {author} {\bibfnamefont {B.}~\bibnamefont
  {Ressel}}, \bibinfo {author} {\bibfnamefont {J.}~\bibnamefont {Slez\'ak}},
  \bibinfo {author} {\bibfnamefont {K.~C.}\ \bibnamefont {Prince}}, \ and\
  \bibinfo {author} {\bibfnamefont {V.}~\bibnamefont {Ch\'ab}},\ }\bibfield
  {title} {\enquote {\bibinfo {title} {{Quantized valence states of the
  Pb/Si(111) mosaic phase}},}\ }\href {\doibase 10.1103/PhysRevB.66.035325}
  {\bibfield  {journal} {\bibinfo  {journal} {Phys. Rev. B}\ }\textbf {\bibinfo
  {volume} {66}},\ \bibinfo {pages} {035325} (\bibinfo {year}
  {2002})}\BibitemShut {NoStop}%
\bibitem [{\citenamefont {Smith}\ \emph {et~al.}(2020)\citenamefont {Smith},
  \citenamefont {Ming}, \citenamefont {Trabada}, \citenamefont {Gonzalez},
  \citenamefont {Soler-Polo}, \citenamefont {Flores}, \citenamefont {Ortega},\
  and\ \citenamefont {Weitering}}]{PhysRevLett.124.097602}%
  \BibitemOpen
  \bibfield  {author} {\bibinfo {author} {\bibfnamefont {T.~S.}\ \bibnamefont
  {Smith}}, \bibinfo {author} {\bibfnamefont {F.}~\bibnamefont {Ming}},
  \bibinfo {author} {\bibfnamefont {D.~G.}\ \bibnamefont {Trabada}}, \bibinfo
  {author} {\bibfnamefont {C.}~\bibnamefont {Gonzalez}}, \bibinfo {author}
  {\bibfnamefont {D.}~\bibnamefont {Soler-Polo}}, \bibinfo {author}
  {\bibfnamefont {F.}~\bibnamefont {Flores}}, \bibinfo {author} {\bibfnamefont
  {J.}~\bibnamefont {Ortega}}, \ and\ \bibinfo {author} {\bibfnamefont {H.~H.}\
  \bibnamefont {Weitering}},\ }\bibfield  {title} {\enquote {\bibinfo {title}
  {{Coupled Sublattice Melting and Charge-Order Transition in Two
  Dimensions}},}\ }\href {\doibase 10.1103/PhysRevLett.124.097602} {\bibfield
  {journal} {\bibinfo  {journal} {Phys. Rev. Lett.}\ }\textbf {\bibinfo
  {volume} {124}},\ \bibinfo {pages} {097602} (\bibinfo {year}
  {2020})}\BibitemShut {NoStop}%
\bibitem [{\citenamefont {{Beyer}}\ \emph {et~al.}(2022)\citenamefont
  {{Beyer}}, \citenamefont {{Hauck}}, \citenamefont {{Klebl}}, \citenamefont
  {{Schwemmer}}, \citenamefont {{Kennes}}, \citenamefont {{Thomale}},
  \citenamefont {{Honerkamp}},\ and\ \citenamefont
  {{Rachel}}}]{2022arXiv221009384B}%
  \BibitemOpen
  \bibfield  {author} {\bibinfo {author} {\bibfnamefont {Jacob}\ \bibnamefont
  {{Beyer}}}, \bibinfo {author} {\bibfnamefont {Jonas~B.}\ \bibnamefont
  {{Hauck}}}, \bibinfo {author} {\bibfnamefont {Lennart}\ \bibnamefont
  {{Klebl}}}, \bibinfo {author} {\bibfnamefont {Tilman}\ \bibnamefont
  {{Schwemmer}}}, \bibinfo {author} {\bibfnamefont {Dante~M.}\ \bibnamefont
  {{Kennes}}}, \bibinfo {author} {\bibfnamefont {Ronny}\ \bibnamefont
  {{Thomale}}}, \bibinfo {author} {\bibfnamefont {Carsten}\ \bibnamefont
  {{Honerkamp}}}, \ and\ \bibinfo {author} {\bibfnamefont {Stephan}\
  \bibnamefont {{Rachel}}},\ }\href@noop {} {\enquote {\bibinfo {title}
  {{Rashba spin-orbit coupling in the square lattice Hubbard model: A
  truncated-unity functional renormalization group study}},}\ }\bibinfo
  {howpublished} {Preprint at \url{https://arxiv.org/abs/2210.09384}} (\bibinfo
  {year} {2022})\BibitemShut {NoStop}%
\bibitem [{\citenamefont {Stepanov}\ \emph {et~al.}(2018)\citenamefont
  {Stepanov}, \citenamefont {Brener}, \citenamefont {Krien}, \citenamefont
  {Harland}, \citenamefont {Lichtenstein},\ and\ \citenamefont
  {Katsnelson}}]{PhysRevLett.121.037204}%
  \BibitemOpen
  \bibfield  {author} {\bibinfo {author} {\bibfnamefont {E.~A.}\ \bibnamefont
  {Stepanov}}, \bibinfo {author} {\bibfnamefont {S.}~\bibnamefont {Brener}},
  \bibinfo {author} {\bibfnamefont {F.}~\bibnamefont {Krien}}, \bibinfo
  {author} {\bibfnamefont {M.}~\bibnamefont {Harland}}, \bibinfo {author}
  {\bibfnamefont {A.~I.}\ \bibnamefont {Lichtenstein}}, \ and\ \bibinfo
  {author} {\bibfnamefont {M.~I.}\ \bibnamefont {Katsnelson}},\ }\bibfield
  {title} {\enquote {\bibinfo {title} {{Effective Heisenberg Model and Exchange
  Interaction for Strongly Correlated Systems}},}\ }\href {\doibase
  10.1103/PhysRevLett.121.037204} {\bibfield  {journal} {\bibinfo  {journal}
  {Phys. Rev. Lett.}\ }\textbf {\bibinfo {volume} {121}},\ \bibinfo {pages}
  {037204} (\bibinfo {year} {2018})}\BibitemShut {NoStop}%
\bibitem [{\citenamefont {Stepanov}\ \emph
  {et~al.}(2019{\natexlab{b}})\citenamefont {Stepanov}, \citenamefont {Huber},
  \citenamefont {Lichtenstein},\ and\ \citenamefont
  {Katsnelson}}]{PhysRevB.99.115124}%
  \BibitemOpen
  \bibfield  {author} {\bibinfo {author} {\bibfnamefont {E.~A.}\ \bibnamefont
  {Stepanov}}, \bibinfo {author} {\bibfnamefont {A.}~\bibnamefont {Huber}},
  \bibinfo {author} {\bibfnamefont {A.~I.}\ \bibnamefont {Lichtenstein}}, \
  and\ \bibinfo {author} {\bibfnamefont {M.~I.}\ \bibnamefont {Katsnelson}},\
  }\bibfield  {title} {\enquote {\bibinfo {title} {{Effective Ising model for
  correlated systems with charge ordering}},}\ }\href {\doibase
  10.1103/PhysRevB.99.115124} {\bibfield  {journal} {\bibinfo  {journal} {Phys.
  Rev. B}\ }\textbf {\bibinfo {volume} {99}},\ \bibinfo {pages} {115124}
  (\bibinfo {year} {2019}{\natexlab{b}})}\BibitemShut {NoStop}%
\bibitem [{\citenamefont {Stepanov}\ \emph
  {et~al.}(2022{\natexlab{b}})\citenamefont {Stepanov}, \citenamefont {Brener},
  \citenamefont {Harkov}, \citenamefont {Katsnelson},\ and\ \citenamefont
  {Lichtenstein}}]{PhysRevB.105.155151}%
  \BibitemOpen
  \bibfield  {author} {\bibinfo {author} {\bibfnamefont {E.~A.}\ \bibnamefont
  {Stepanov}}, \bibinfo {author} {\bibfnamefont {S.}~\bibnamefont {Brener}},
  \bibinfo {author} {\bibfnamefont {V.}~\bibnamefont {Harkov}}, \bibinfo
  {author} {\bibfnamefont {M.~I.}\ \bibnamefont {Katsnelson}}, \ and\ \bibinfo
  {author} {\bibfnamefont {A.~I.}\ \bibnamefont {Lichtenstein}},\ }\bibfield
  {title} {\enquote {\bibinfo {title} {{Spin dynamics of itinerant electrons:
  Local magnetic moment formation and Berry phase}},}\ }\href {\doibase
  10.1103/PhysRevB.105.155151} {\bibfield  {journal} {\bibinfo  {journal}
  {Phys. Rev. B}\ }\textbf {\bibinfo {volume} {105}},\ \bibinfo {pages}
  {155151} (\bibinfo {year} {2022}{\natexlab{b}})}\BibitemShut {NoStop}%
\bibitem [{\citenamefont {Ramazanov}(2011)}]{Ramazanov2011}%
  \BibitemOpen
  \bibfield  {author} {\bibinfo {author} {\bibfnamefont {M.~K.}\ \bibnamefont
  {Ramazanov}},\ }\bibfield  {title} {\enquote {\bibinfo {title} {Phase
  transitions in the antiferromagnetic heisenberg model on a layered triangular
  lattice with the next-nearest neighbor interactions},}\ }\href {\doibase
  10.1134/S0021364011160156} {\bibfield  {journal} {\bibinfo  {journal} {JETP
  Letters}\ }\textbf {\bibinfo {volume} {94}},\ \bibinfo {pages} {311}
  (\bibinfo {year} {2011})}\BibitemShut {NoStop}%
\bibitem [{\citenamefont {Serrate}\ \emph {et~al.}(2010)\citenamefont
  {Serrate}, \citenamefont {Ferriani}, \citenamefont {Yoshida}, \citenamefont
  {Hla}, \citenamefont {Menzel}, \citenamefont {von Bergmann}, \citenamefont
  {Heinze}, \citenamefont {Kubetzka},\ and\ \citenamefont
  {Wiesendanger}}]{Serrate2010}%
  \BibitemOpen
  \bibfield  {author} {\bibinfo {author} {\bibfnamefont {David}\ \bibnamefont
  {Serrate}}, \bibinfo {author} {\bibfnamefont {Paolo}\ \bibnamefont
  {Ferriani}}, \bibinfo {author} {\bibfnamefont {Yasuo}\ \bibnamefont
  {Yoshida}}, \bibinfo {author} {\bibfnamefont {Saw-Wai}\ \bibnamefont {Hla}},
  \bibinfo {author} {\bibfnamefont {Matthias}\ \bibnamefont {Menzel}}, \bibinfo
  {author} {\bibfnamefont {Kirsten}\ \bibnamefont {von Bergmann}}, \bibinfo
  {author} {\bibfnamefont {Stefan}\ \bibnamefont {Heinze}}, \bibinfo {author}
  {\bibfnamefont {Andre}\ \bibnamefont {Kubetzka}}, \ and\ \bibinfo {author}
  {\bibfnamefont {Roland}\ \bibnamefont {Wiesendanger}},\ }\bibfield  {title}
  {\enquote {\bibinfo {title} {{Imaging and manipulating the spin direction of
  individual atoms}},}\ }\href {\doibase 10.1038/nnano.2010.64} {\bibfield
  {journal} {\bibinfo  {journal} {Nat. Nanotechnol.}\ }\textbf {\bibinfo
  {volume} {5}},\ \bibinfo {pages} {350--353} (\bibinfo {year}
  {2010})}\BibitemShut {NoStop}%
\bibitem [{\citenamefont {Nair}\ \emph {et~al.}(2023)\citenamefont {Nair},
  \citenamefont {Palacio}, \citenamefont {Mascaraque}, \citenamefont {Michel},
  \citenamefont {Taleb-Ibrahimi}, \citenamefont {Tejeda}, \citenamefont
  {Gonz\'alez}, \citenamefont {Mart\'{\i}n-Rodero}, \citenamefont {Ortega},\
  and\ \citenamefont {Flores}}]{PhysRevB.107.045303}%
  \BibitemOpen
  \bibfield  {author} {\bibinfo {author} {\bibfnamefont {M.~N.}\ \bibnamefont
  {Nair}}, \bibinfo {author} {\bibfnamefont {I.}~\bibnamefont {Palacio}},
  \bibinfo {author} {\bibfnamefont {A.}~\bibnamefont {Mascaraque}}, \bibinfo
  {author} {\bibfnamefont {E.~G.}\ \bibnamefont {Michel}}, \bibinfo {author}
  {\bibfnamefont {A.}~\bibnamefont {Taleb-Ibrahimi}}, \bibinfo {author}
  {\bibfnamefont {A.}~\bibnamefont {Tejeda}}, \bibinfo {author} {\bibfnamefont
  {C.}~\bibnamefont {Gonz\'alez}}, \bibinfo {author} {\bibfnamefont
  {A.}~\bibnamefont {Mart\'{\i}n-Rodero}}, \bibinfo {author} {\bibfnamefont
  {J.}~\bibnamefont {Ortega}}, \ and\ \bibinfo {author} {\bibfnamefont
  {F.}~\bibnamefont {Flores}},\ }\bibfield  {title} {\enquote {\bibinfo {title}
  {Giant electron-phonon interaction for a prototypical semiconductor
  interface:
  $\mathrm{Sn}/\mathrm{Ge}(111)\text{\ensuremath{-}}(3\ifmmode\times\else\texttimes\fi{}3)$},}\
  }\href {\doibase 10.1103/PhysRevB.107.045303} {\bibfield  {journal} {\bibinfo
   {journal} {Phys. Rev. B}\ }\textbf {\bibinfo {volume} {107}},\ \bibinfo
  {pages} {045303} (\bibinfo {year} {2023})}\BibitemShut {NoStop}%
\bibitem [{\citenamefont {Wu}\ \emph {et~al.}(2008)\citenamefont {Wu},
  \citenamefont {Phillips},\ and\ \citenamefont
  {Castro~Neto}}]{PhysRevLett.101.126401}%
  \BibitemOpen
  \bibfield  {author} {\bibinfo {author} {\bibfnamefont {Jiansheng}\
  \bibnamefont {Wu}}, \bibinfo {author} {\bibfnamefont {Philip}\ \bibnamefont
  {Phillips}}, \ and\ \bibinfo {author} {\bibfnamefont {A.~H.}\ \bibnamefont
  {Castro~Neto}},\ }\bibfield  {title} {\enquote {\bibinfo {title} {{Theory of
  the Magnetic Moment in Iron Pnictides}},}\ }\href {\doibase
  10.1103/PhysRevLett.101.126401} {\bibfield  {journal} {\bibinfo  {journal}
  {Phys. Rev. Lett.}\ }\textbf {\bibinfo {volume} {101}},\ \bibinfo {pages}
  {126401} (\bibinfo {year} {2008})}\BibitemShut {NoStop}%
\bibitem [{\citenamefont {Profeta}\ and\ \citenamefont
  {Tosatti}(2007{\natexlab{b}})}]{Profeta2007}%
  \BibitemOpen
  \bibfield  {author} {\bibinfo {author} {\bibfnamefont {G.}~\bibnamefont
  {Profeta}}\ and\ \bibinfo {author} {\bibfnamefont {E.}~\bibnamefont
  {Tosatti}},\ }\bibfield  {title} {\enquote {\bibinfo {title} {Triangular
  mott-hubbard insulator phases of $\mathrm{Sn}/\mathrm{Si}(111)$ and
  $\mathrm{Sn}/\mathrm{Ge}(111)$ surfaces},}\ }\href {\doibase
  10.1103/PhysRevLett.98.086401} {\bibfield  {journal} {\bibinfo  {journal}
  {Phys. Rev. Lett.}\ }\textbf {\bibinfo {volume} {98}},\ \bibinfo {pages}
  {086401} (\bibinfo {year} {2007}{\natexlab{b}})}\BibitemShut {NoStop}%
\bibitem [{\citenamefont {Blaha}\ \emph {et~al.}(2018)\citenamefont {Blaha},
  \citenamefont {Schwarz}, \citenamefont {Madsen}, \citenamefont {Kvasnicka},
  \citenamefont {Luitz}, \citenamefont {Laskowski}, \citenamefont {Tran},\ and\
  \citenamefont {Marks}}]{wien2k}%
  \BibitemOpen
  \bibfield  {author} {\bibinfo {author} {\bibfnamefont {P}~\bibnamefont
  {Blaha}}, \bibinfo {author} {\bibfnamefont {K}~\bibnamefont {Schwarz}},
  \bibinfo {author} {\bibfnamefont {G}~\bibnamefont {Madsen}}, \bibinfo
  {author} {\bibfnamefont {D}~\bibnamefont {Kvasnicka}}, \bibinfo {author}
  {\bibfnamefont {J}~\bibnamefont {Luitz}}, \bibinfo {author} {\bibfnamefont
  {R}~\bibnamefont {Laskowski}}, \bibinfo {author} {\bibfnamefont
  {F}~\bibnamefont {Tran}}, \ and\ \bibinfo {author} {\bibfnamefont {L.~D.}\
  \bibnamefont {Marks}},\ }\href@noop {} {\emph {\bibinfo {title} {WIEN2k, An
  augmented Plane Wave + Local Orbitals Program for Calculating Crystal
  Properties}}}\ (\bibinfo  {publisher} {Karlheinz Schwarz, Techn. Universität
  Wien, Austria,ISBN 3-9501031-1-2},\ \bibinfo {year} {2018})\BibitemShut
  {NoStop}%
\bibitem [{\citenamefont {Blaha}\ \emph {et~al.}(2020)\citenamefont {Blaha},
  \citenamefont {Schwarz}, \citenamefont {Tran}, \citenamefont {Laskowski},
  \citenamefont {Madsen},\ and\ \citenamefont {Marks}}]{wien2k2020}%
  \BibitemOpen
  \bibfield  {author} {\bibinfo {author} {\bibfnamefont {Peter}\ \bibnamefont
  {Blaha}}, \bibinfo {author} {\bibfnamefont {Karlheinz}\ \bibnamefont
  {Schwarz}}, \bibinfo {author} {\bibfnamefont {Fabien}\ \bibnamefont {Tran}},
  \bibinfo {author} {\bibfnamefont {Robert}\ \bibnamefont {Laskowski}},
  \bibinfo {author} {\bibfnamefont {Georg K.~H.}\ \bibnamefont {Madsen}}, \
  and\ \bibinfo {author} {\bibfnamefont {Laurence~D.}\ \bibnamefont {Marks}},\
  }\bibfield  {title} {\enquote {\bibinfo {title} {{WIEN2k: An {APW}+lo program
  for calculating the properties of solids}},}\ }\href {\doibase
  10.1063/1.5143061} {\bibfield  {journal} {\bibinfo  {journal} {Chem. Phys.}\
  }\textbf {\bibinfo {volume} {152}},\ \bibinfo {pages} {074101} (\bibinfo
  {year} {2020})}\BibitemShut {NoStop}%
\bibitem [{\citenamefont {Georges}\ \emph {et~al.}(1996)\citenamefont
  {Georges}, \citenamefont {Kotliar}, \citenamefont {Krauth},\ and\
  \citenamefont {Rozenberg}}]{RevModPhys.68.13}%
  \BibitemOpen
  \bibfield  {author} {\bibinfo {author} {\bibfnamefont {Antoine}\ \bibnamefont
  {Georges}}, \bibinfo {author} {\bibfnamefont {Gabriel}\ \bibnamefont
  {Kotliar}}, \bibinfo {author} {\bibfnamefont {Werner}\ \bibnamefont
  {Krauth}}, \ and\ \bibinfo {author} {\bibfnamefont {Marcelo~J.}\ \bibnamefont
  {Rozenberg}},\ }\bibfield  {title} {\enquote {\bibinfo {title} {Dynamical
  mean-field theory of strongly correlated fermion systems and the limit of
  infinite dimensions},}\ }\href {\doibase 10.1103/RevModPhys.68.13} {\bibfield
   {journal} {\bibinfo  {journal} {Rev. Mod. Phys.}\ }\textbf {\bibinfo
  {volume} {68}},\ \bibinfo {pages} {13--125} (\bibinfo {year}
  {1996})}\BibitemShut {NoStop}%
\bibitem [{\citenamefont {Wallerberger}\ \emph {et~al.}(2019)\citenamefont
  {Wallerberger}, \citenamefont {Hausoel}, \citenamefont {Gunacker},
  \citenamefont {Kowalski}, \citenamefont {Parragh}, \citenamefont {Goth},
  \citenamefont {Held},\ and\ \citenamefont
  {Sangiovanni}}]{WALLERBERGER2019388}%
  \BibitemOpen
  \bibfield  {author} {\bibinfo {author} {\bibfnamefont {Markus}\ \bibnamefont
  {Wallerberger}}, \bibinfo {author} {\bibfnamefont {Andreas}\ \bibnamefont
  {Hausoel}}, \bibinfo {author} {\bibfnamefont {Patrik}\ \bibnamefont
  {Gunacker}}, \bibinfo {author} {\bibfnamefont {Alexander}\ \bibnamefont
  {Kowalski}}, \bibinfo {author} {\bibfnamefont {Nicolaus}\ \bibnamefont
  {Parragh}}, \bibinfo {author} {\bibfnamefont {Florian}\ \bibnamefont {Goth}},
  \bibinfo {author} {\bibfnamefont {Karsten}\ \bibnamefont {Held}}, \ and\
  \bibinfo {author} {\bibfnamefont {Giorgio}\ \bibnamefont {Sangiovanni}},\
  }\bibfield  {title} {\enquote {\bibinfo {title} {{w2dynamics: Local one- and
  two-particle quantities from dynamical mean field theory}},}\ }\href
  {\doibase https://doi.org/10.1016/j.cpc.2018.09.007} {\bibfield  {journal}
  {\bibinfo  {journal} {Comput. Phys. Commun.}\ }\textbf {\bibinfo {volume}
  {235}},\ \bibinfo {pages} {388--399} (\bibinfo {year} {2019})}\BibitemShut
  {NoStop}%
\end{thebibliography}%

\clearpage
\appendix

\end{document}